\documentclass[aps,prb,preprint,superscriptaddress]{revtex4-2}
\usepackage{gensymb}
\usepackage{amssymb}
\usepackage{graphicx}
\usepackage{booktabs}
\usepackage{multirow}
\usepackage{chemformula}
\usepackage{appendix}
\begin{document}
\title{Low-temperature magnetic behaviour on the triangular lattice in hexagonal \ch{Ba3Tb(BO3)3}}
\author{Nicola D.~Kelly}
\email[]{ne281@cam.ac.uk}
\affiliation{Cavendish Laboratory, University of Cambridge, J J Thomson Avenue, Cambridge, CB3 0HE, UK}
\affiliation{Jesus College, University of Cambridge, Jesus Lane, CB5 8BL, UK}
\author{M.~Duc Le}
\affiliation{ISIS Neutron and Muon Source, Rutherford Appleton Laboratory, Didcot, OX11 0QX, UK}
\author{Denis Sheptyakov}
\affiliation{Laboratory for Neutron Scattering and Imaging, Paul Scherrer Institut, Forschungsstrasse 111, Villigen, CH-5232, Switzerland}
\author{Camilla Tacconis}
\affiliation{Cavendish Laboratory, University of Cambridge, J J Thomson Avenue, Cambridge, CB3 0HE, UK}
\author{Cheng Liu}
\affiliation{Cavendish Laboratory, University of Cambridge, J J Thomson Avenue, Cambridge, CB3 0HE, UK}
\author{Gavin B.~G.~Stenning}
\affiliation{ISIS Neutron and Muon Source, Rutherford Appleton Laboratory, Didcot, OX11 0QX, UK}
\author{Peter J.~Baker}
\affiliation{ISIS Neutron and Muon Source, Rutherford Appleton Laboratory, Didcot, OX11 0QX, UK}
\author{Si\^{a}n E.~Dutton}
\email[]{sed33@cam.ac.uk}
\affiliation{Cavendish Laboratory, University of Cambridge, J J Thomson Avenue, Cambridge, CB3 0HE, UK}
\date{\today}
\begin{abstract}
The hexagonal polymorph of \ch{Ba3Tb(BO3)3} contains Tb$^{3+}$ ions on a quasi-2D triangular lattice, resulting in geometric magnetic frustration. Powder samples of \ch{Ba3Tb(BO3)3} have been investigated using specific heat, powder neutron diffraction (PND), inelastic neutron scattering (INS) and muon-spin relaxation spectroscopy ($\mu$SR). No long-range magnetic ordering is observed down to the lowest measured temperatures of 75~mK in PND and specific heat data and 1.5~K in the $\mu$SR data. Modelling the INS spectrum using a point charge model suggests that the ground state is a singlet with a low-lying doublet on each of the two crystallographically independent Tb$^{3+}$ sites and that both the Tb ions display weak \textit{XY} single-ion anisotropy.
\end{abstract}
\maketitle

\section{Introduction}
Ensembles of magnetic spins are frequently discussed using the analogy of the three states of matter: gas (paramagnet), liquid, and solid (ordered ferromagnet or antiferromagnet). In the case of a spin liquid the spins are expected to display some short-range correlations but no long-range order (LRO), as molecules do in a real liquid; at some low temperature $T_f$ the system often `freezes' into a spin glass, where the spins are stationary and orientationally disordered across the material. Quantum spin liquids (QSLs) are an exotic state of matter where the spins exhibit short-range correlations, yet resist LRO and remain very strongly fluctuating down to very low temperatures. Typical features for QSL candidates include quantum spins ($S=\frac{1}{2}$, e.g.~Cu$^{2+}$, Yb$^{3+}$) combined with magnetic frustration and/or low dimensionality \cite{Vojta2018}. Many candidate materials combine all three of these features, e.g.~the layered mineral herbertsmithite, \ch{ZnCu3(OH)6Cl2} \cite{Shores2005}. However, Cu$^{2+}$/Zn$^{2+}$ site mixing produces defects in the perfect kagome lattice in this case \cite{Vasiliev2018}. Another well-known QSL candidate is \ch{YbMgGaO4} which contains a perfect triangular lattice of Yb$^{3+}$ \cite{Li2015}. However, \ch{YbMgGaO4} suffers from non-magnetic Mg$^{2+}$/Ga$^{3+}$ site disorder, which is harmful to the formation of a true spin liquid state through localization of spinons \cite{Xu2016}, and the exact nature of the ground state is still the subject of debate \cite{Li2016,Paddison2017,Li2017,Ma2018}.

By contrast with the examples of herbertsmithite and \ch{YbMgGaO4}, the crystal structure of the layered triangular lattice compounds \ch{Ba3\textit{Ln}(BO3)3} (\textit{Ln}~= Y, Dy--Lu) contains structurally distinct \textit{Ln}$^{3+}$ triangular layers with no Ba$^{2+}$ substitution or disorder within the non-magnetic layers \cite{Khamaganova1999,Simura2017,Gao2018}. None of these materials shows long-range magnetic order above 2~K in bulk susceptibility measurements \cite{Gao2018}. The $S=\frac{1}{2}$ compound \ch{Ba3Yb(BO3)3} has additionally been investigated using nuclear magnetic resonance (NMR) spectroscopy down to $T=0.26$~K, showing a lack of spin ordering or freezing and the presence of strong quantum spin fluctuations \cite{Zeng2020}. These observations are supported by a recent muon-spin relaxation spectroscopy ($\mu$SR) study which concludes that \ch{Ba3Yb(BO3)3} remains paramagnetic down to $T=0.28$~K with no long-range ordering or spin freezing \cite{Cho2021}. However, Bag \textit{et al.}~suggested that \ch{Ba3Yb(BO3)3} is an example of a pure dipolar system, in which the long-range dipole-dipole interaction between $S=\frac{1}{2}$ Yb$^{3+}$ spins dominates over the much smaller or negligible superexchange interaction \cite{Bag2021}. Ennis \textit{et al}.~also investigated the $S=\frac{3}{2}$ analogue \ch{Ba3Er(BO3)3} in both powder and single-crystal form and observed two-sublattice exchange interactions, with a proposed antiferromagnetic ordering transition at 100~mK \cite{Ennis2024}.

\begin{figure}[htbp]
\centering
\includegraphics[width=8cm]{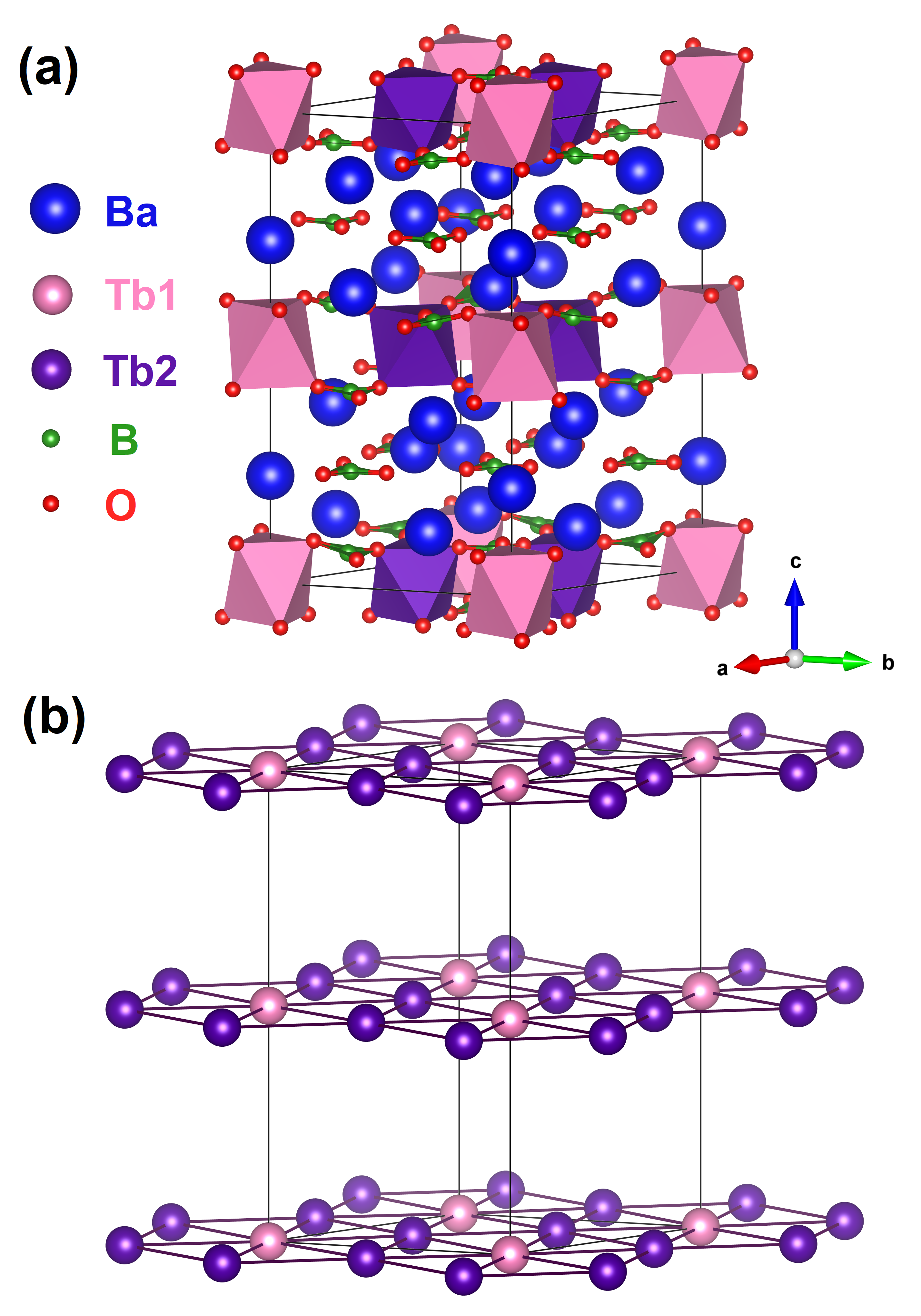}
\caption{(a) Crystal structure of hexagonal \ch{Ba3Tb(BO3)3}, space group $P6_3cm$. Red spheres are O, green are B, blue are Ba and pink and purple are Tb sites 1 and 2 respectively. (b) shows the arrangement of Tb$^{3+}$ ions only. Figure produced using \textsc{Vesta} \cite{Momma2011}.}
\label{fig:batbstructure}
\end{figure}

We previously reported the synthesis of a new low-temperature hexagonal phase of \ch{Ba3Tb(BO3)3} (Fig.~\ref{fig:batbstructure}), isostructural with the layered compounds \ch{Ba3\textit{Ln}(BO3)3} (\textit{Ln}~= Y, Dy--Lu). Similar to its heavier analogues, bulk magnetic measurements on \ch{Ba3Tb(BO3)3} revealed antiferromagnetic interactions with no 3D magnetic ordering transition above 2~K, although there was evidence for possible short-range magnetic ordering and/or field-induced ordering in the $\chi'(T)$ and $M(H)$ data \cite{Kelly2020b}. We here present subsequent measurements of low-temperature specific heat, elastic and inelastic neutron scattering, and muon-spin relaxation of polycrystalline \ch{Ba3Tb(BO3)3}. The lack of magnetic Bragg or diffuse scattering down to low temperatures (nominally 75~mK) indicates an absence of three-dimensional LRO with a large frustration parameter \cite{Ramirez1994} of $f=|\theta_\mathrm{CW}/T_\mathrm{N}| \gtrapprox 95$. Muon-spin spectroscopy down to 1.5~K shows neither sinusoidal oscillations indicating LRO nor a ``1/3 tail'' suggesting a spin glass. A low-$T$ upturn in the specific heat, which otherwise shows no signatures of LRO, is attributed to a nuclear Schottky anomaly of the Tb$^{3+}$ ion. Finally, fitting of the inelastic neutron spectrum to a point charge model suggests that the ground state of \ch{Ba3Tb(BO3)3} is a singlet on both the Tb sites.

\section{Experimental Methods}
\textit{Synthesis}. Polycrystalline \ch{Ba3Tb(BO3)3} in the low-temperature hexagonal ($P6_3cm$) phase was synthesised by a solid-state procedure on a 0.5--3.0~g scale as reported previously \cite{Kelly2020b}. \ch{BaCO3} (Alfa Aesar, 99.997~\%), \ch{Tb4O7} (Alfa Aesar, 99.998~\%) and \ch{H3BO3} (Alfa Aesar, 99.999~\%) were ground together in stoichiometric amounts, preheated at 500~\degree C for 12~h, then pressed into a pellet and heated to 875~\degree C for several days with regrinding every 12~h until phase pure. Samples for neutron scattering experiments were synthesised using $^{11}$B-enriched boric acid (99 atom~\%, Sigma) to avoid the strong neutron absorption of the $^{10}$B isotope. Phase purity was checked using powder X-ray diffraction (PXRD) on a Bruker D8 diffractometer with Cu K$\alpha$ radiation, $\lambda = 1.541$~\AA.

\textit{Structural characterisation}. We used powder neutron diffraction (PND) to confirm the crystal structure of hexagonal \ch{Ba3Tb(BO3)3} at room temperature, particularly the positions of lighter boron and oxygen atoms, to which X-rays are less sensitive. The sample was packed into a vanadium canister and data collected at $T=300$~K in the range $3.90\leq2\theta$~(\degree) $\leq164.85$, step size 0.05\degree, wavelength $\lambda\approx1.494$~\AA, on the HRPT beamline \cite{Fischer2000} at the Paul Scherrer Institute, Switzerland. PXRD ($\lambda=$~Cu K$\alpha$) had previously been carried out on the same sample to obtain lattice parameters and atomic positions for Tb and Ba atoms using Rietveld refinement \cite{Rietveld1969} in \textsc{Topas} \cite{Coelho2018}; the starting positions for B and O atoms were taken from the literature report for \ch{Ba3Yb(BO3)3} \cite{Khamaganova1999}. Initially, these lattice parameters and atomic positions were fixed in order to refine the neutron wavelength. Subsequently the wavelength was fixed at the refined value of 1.49426(2)~\AA\ and the lattice parameters and all atomic coordinates, including B and O atoms, were refined. The $B_\mathrm{iso}$ values were refined with the constraint $B_\mathrm{iso}\geq0$ for each atom. Additional PND patterns were collected on the same sample in a cryostat at 30, 5 and 1.5~K, and in a dilution refrigerator at 5~K and 75~mK, on the same beamline HRPT at PSI. The sub-ambient measurements utilised a longer wavelength $\lambda=2.45$~\AA\ to optimise the $Q$-range for magnetic neutron scattering. A copper container was chosen for its high thermal conductivity, giving rise to large nuclear Bragg peaks at approximately $2\theta=72$\degree, 86\degree, 148\degree. The container was filled to an overpressure of helium (10 bar) in order to enable efficient heat transfer between the container and all the grains of powder. For all Rietveld refinements, the background was modelled with a 12-coefficient Chebyshev polynomial and the peak shape was modelled with a modified Thompson-Cox-Hastings pseudo-Voigt function with axial divergence asymmetry \cite{Young1993}.

\textit{Inelastic neutron scattering}. Time-of-flight INS was carried out on the MARI beamline at the ISIS Neutron and Muon Source \cite{Kelly2020MARI} on a 2~g powder sample of \ch{Ba3Tb(^{11}BO3)3} in a aluminium sample canister. The temperature was controlled using a helium-flow insert in a closed cycle refrigerator. The incident neutron energies were $E_\mathrm{i}=100$ (23, 10)~meV (Gd chopper at 200~Hz) and 180 (30)~meV (chopper at 400~Hz) and the temperatures for data collection were $T=1.5$, 10, 25 and 50~K.

\textit{Specific heat}. Approximately 40~mg of sample was ground with an equal mass of Ag powder (Alfa Aesar, 99.99\%, --635 mesh) using a pestle and mortar, then pressed into a pellet of diameter 5~mm. Portions of this pellet weighing 5--20~mg were attached to the sample holder using Apiezon N grease. The zero-field heat capacity at low temperatures was measured at $T\geq2$~K using the standard heat capacity option and $T\geq0.4$~K using the He3 option on a Quantum Design Physical Properties Measurement System (PPMS). Finally, the heat capacity was measured in the range $0.05\leq T$(K) $\leq 4$ using a dilution insert to the PPMS at ISIS. The Ag lattice contribution to the total heat capacity was modelled using a $T^3$ dependence, $C_\mathrm{p}=\gamma T+AT^3$, with coefficients taken from the literature \cite{Arblaster2015}. The \ch{Ba3Tb(BO3)3} lattice contribution was modelled using a $T^3$ (Debye) dependence with $\theta_\mathrm{D}=255$~K \cite{Gopal1966}.

\textit{Muon-spin relaxation spectroscopy}. $\mu$SR was carried out on the EMU beamline at ISIS \cite{Kelly2021EMU}. The polycrystalline sample ($m\approx1$~g) was placed in an envelope of silver foil before being attached to the sample stick and inserted into the helium cryostat ($T_\mathrm{min}=1.5$~K). Data were collected in zero field (ZF) and in an applied longitudinal field (LF) of 200~G at temperatures between 1.5 and 300~K. Additional LF scans from 0--4000~G were collected at $T=1.5$, 10, 50 and 100~K. Data analysis was carried out using the program \textsc{Mantid} \cite{Arnold2014}.

\section{Results and Discussion}

\subsection{Powder neutron diffraction}
\begin{figure}[htbp]
\centering
\includegraphics[width=15cm]{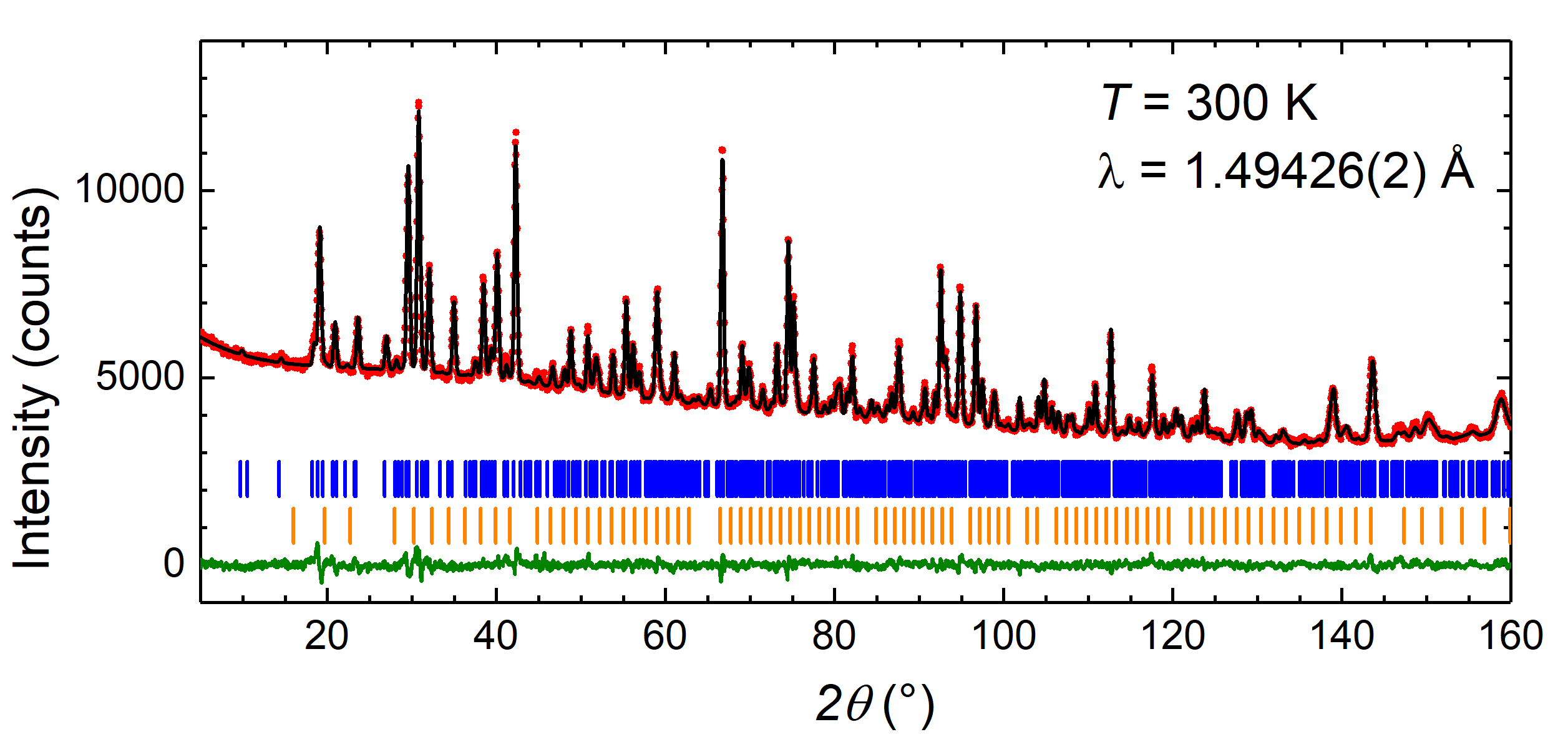}
\caption{Rietveld refinement of PND data for \ch{Ba3Tb(BO3)3} at 300~K. Red dots -- observed; black line -- calculated; green line -- difference; tick marks -- Bragg reflection positions for \ch{Ba3Tb(BO3)3} (blue), \ch{Tb2O3} (orange; 4.2(4)~wt~\%). For clarity, the observed and calculated intensities have each been offset by 1000 counts in the positive $y$-direction.}
\label{fig:rtpndbatb}
\end{figure}

In order to confirm the crystal structure, particularly the positions of the lighter boron and oxygen ions, an $^{11}$B-enriched sample of \ch{Ba3Tb(BO3)3} was studied using powder neutron diffraction (PND) at room temperature. The refined structural parameters are given in Table~\ref{table:batbatomsn} and are in good agreement with previous X-ray diffraction studies \cite{Kelly2020b,Khamaganova1999}. Bond valence sum parameters $V$, calculated according to $V_i=\sum_is_{ij}=\sum_ie^{\frac{R_0-R_{ij}}{B}}$, \cite{Brown1985} agree reasonably well with the expected valences for all atoms. The structural refinement of normalised PND data at 300~K is shown in Fig.~\ref{fig:rtpndbatb}. The crystal structure of \ch{Ba3Tb(BO3)3} (Fig.~\ref{fig:batbstructure}) is built from almost perfect triangular layers of \ch{TbO6} distorted octahedra; there are two crystallographically distinct Tb$^{3+}$ sites, Tb1 and Tb2, with multiplicity 2 and 4 respectively. The \ch{TbO6} polyhedra are linked together within their layers by corner-sharing with trigonal planar \ch{BO3} groups. Between the terbium-containing layers there are layers containing edge- and face-sharing \ch{BaO9} polyhedra and additional \ch{BO3} groups. A small amount (4.2(4)~wt~\%) of \ch{Tb2O3} is observed by PND in addition to the main phase. This impurity was present in the larger, $^{11}B$-enriched sample used for PND and could not be eliminated before the scheduled beamtime. Smaller samples of \ch{Ba3Tb(BO3)3} used for previous susceptibility measurements \cite{Kelly2020b} and heat capacity measurements in this work were phase-pure by PXRD. A table of samples used is provided in the Supplemental Material \cite{supplemental}.

\begin{table}[htbp]
\centering
\caption{Atomic positions for \ch{Ba3Tb(BO3)3} ($P6_3cm$) from powder neutron diffraction at 300~K, $\lambda=1.494$~\AA. The refined lattice parameters are: $a=b=$~9.44342(9)~\AA; $c$ = 17.7198(3)~\AA; $V=1368.51(4)$~\AA$^3$ ($Z$~= 6). The goodness-of-fit parameter $\chi^2=1.85$ and $R_{wp}=2.25$~\%.}
\label{table:batbatomsn}
\resizebox{\columnwidth}{!}{
\begin{ruledtabular}
\begin{tabular}{c c c c c c c}
Atom & Site & $x$ & $y$ & $z$ & $B_\mathrm{iso}$ (\AA$^2$) & BVS \\
\midrule
Tb1 & $2a\,(0,0,z)$ & 0 & 0 & 0 & 0.6(4) & $+2.56$ \\  
Tb2 & $4b\,(\frac{2}{3},\frac{1}{3},z)$ & $\frac{2}{3}$ & $\frac{1}{3}$ & 0.0021(10) & 0.41(19) & $+3.44$ \\ 
\midrule
Ba1 & $2a\,(0,0,z)$ & 0 & 0 & 0.2265(17) & 0.4(4) & $+2.18$ \\  
Ba2 & $4b\,(\frac{2}{3},\frac{1}{3},z)$ & $\frac{2}{3}$ & $\frac{1}{3}$ & 0.2735(11) & 0.6(2) & $+2.02$ \\  
Ba3 & $6c\,(x,0,z)$ & 0.660(2) & 0 & 0.1270(11) & 0.04(18) & $+2.07$ \\  
Ba4 & $6c\,(x,0,z)$ & 0.675(3) & 0 & 0.3651(12) & 1.2(3) & $+2.12$ \\  
\midrule
B1 & $6c\,(x,0,z)$ & 0.6761(17) & 0 & 0.5737(10) & 0.71(14) & $+2.67$\\  
B2 & $6c\,(x,0,z)$ & 0.667(2) & 0 & 0.7462(11) & 0.41(5) & $+2.84$ \\  
B3 & $6c\,(x,0,z)$ & 0.6651(15) & 0 & 0.9163(10) & 0.13(12) & $+3.10$ \\  
\midrule
O1 & $6c\,(x,0,z)$ & 0.8216(12) & 0 & 0.5951(10) & 0.43(16) & $-1.91$\\
O2 & $12d\,(x,y,z)$ & 0.3286(19) & 0.4716(17) & 0.5709(10) & 1.40(18) & $-2.00$\\
O3 & $12d\,(x,y,z)$ & 0.1867(16) & 0.336(2) & 0.7488(10) & 0.46(16) & $-2.07$\\
O4 & $6c\,(x,0,z)$ & 0.521(3) & 0 & 0.7466(13) & 1.6(4) & $-2.08$\\
O5 & $12d\,(x,y,z)$ & 0.3375(18) & 0.4834(14) & 0.9165(10) & 0.50(11) & $-1.89$ \\
O6 & $6c\,(x,0,z)$ & 0.808(3) & 0 & 0.9216(11) & 1.3(3) & $-2.12$ \\
\end{tabular}
\end{ruledtabular}
}
\end{table}

\begin{figure}[htbp]
\centering
\includegraphics[width=15cm]{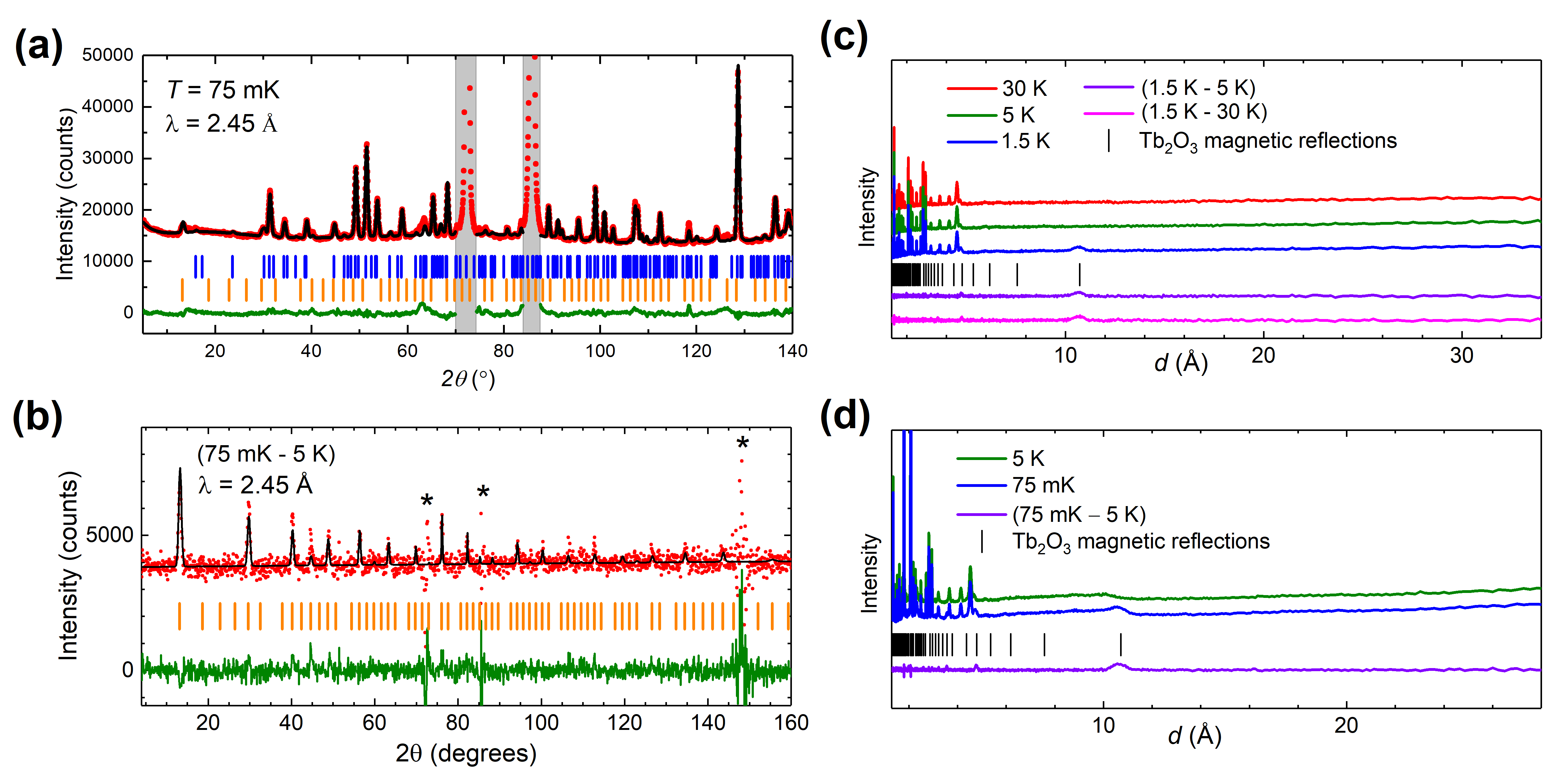}
\caption{HRPT powder neutron diffraction data collected at low temperatures with $\lambda=2.45$~\AA. (a) Rietveld refinement of PND data for \ch{Ba3Tb(BO3)3} at 75~mK. Red dots -- observed; black line -- calculated; green line -- difference; tick marks -- Bragg reflection positions for \ch{Ba3Tb(BO3)3} (blue), \ch{Tb2O3} (orange). For clarity, the observed and calculated intensities have each been offset by 5000 counts in the positive $y$-direction. The grey shaded areas indicate excluded regions in the Rietveld fit due to the Cu sample container. (b) Rietveld fit to dilution fridge difference profile (75~mK -- 5~K). * -- Cu sample container reflections. (c) Data collected in the cryostat ($T\geq1.5$~K) and two difference profiles showing the magnetic peaks from \ch{Tb2O3}. (d) Data collected in the dilution fridge ($T\geq75$~mK) and the difference profile.}
\label{fig:batbpndx4}
\end{figure}

Powder neutron diffraction patterns were also collected at $T=30$, 5, and 1.5~K using a cryostat sample environment and a longer wavelength (2.45~\AA) to optimise the $Q$-range for magnetic scattering. Upon cooling in the neutron beam from 300 to 30~K, the Ba1 and Ba2 atoms changed position in the z-direction, causing a significant change to the arrangement of coordination polyhedra (see Supplemental Material \cite{supplemental}). However, there were no changes to the arrangement of Tb$^{3+}$ ions in the 2D triangular lattice. The crystal structure, including the Ba$^{2+}$ ions, remained the same on further cooling from 30~K to 1.5~K and there were no significant changes to the boron and oxygen positions across the entire temperature range studied. The PND pattern collected at 1.5~K displayed additional Bragg peaks which could all be indexed to the \ch{Tb2O3} impurity phase, which orders antiferromagnetically at $T_\mathrm{N}=2.4$~K \cite{MacChesney1966,Ayant1971,Hill1986}. Details of the magnetic structure of \ch{Tb2O3} along with the structural refinements of PND data for \ch{Ba3Tb(BO3)3} at 30, 5 and 1.5~K and a table of lattice parameters and atomic positions are given in the Supplemental Material \cite{supplemental}.

Additional PND patterns were collected on the same sample in a dilution refrigerator at 5~K and 75~mK, again using a wavelength of 2.45~\AA. At the base temperature, all peaks could be indexed to the \ch{Ba3Tb(BO3)3} nuclear cell, \ch{Tb2O3} nuclear and magnetic cells, or the Cu sample container, Fig.~\ref{fig:batbpndx4}(a,b). No substantial diffuse scattering was observed between 30~K and 75~mK across the range of $d$-spacings probed in this experiment, in either the cryostat or dilution refrigerator environments, Fig.~\ref{fig:batbpndx4}(c,d). Fig.~\ref{fig:batbpndx4}(b) shows the PND difference profile (75~mK -- 5~K) collected in the dilution fridge, fitted to the magnetic model for \ch{Tb2O3} described above. Using this model all of the peaks in the difference profiles can be accounted for, which implies that \ch{Ba3Tb(BO3)3} does not exhibit magnetic ordering down to a nominal temperature of 75~mK.

\subsection{Inelastic neutron scattering} \label{section:ins}

\begin{figure}[htbp] 
\centering
\includegraphics[width=8cm]{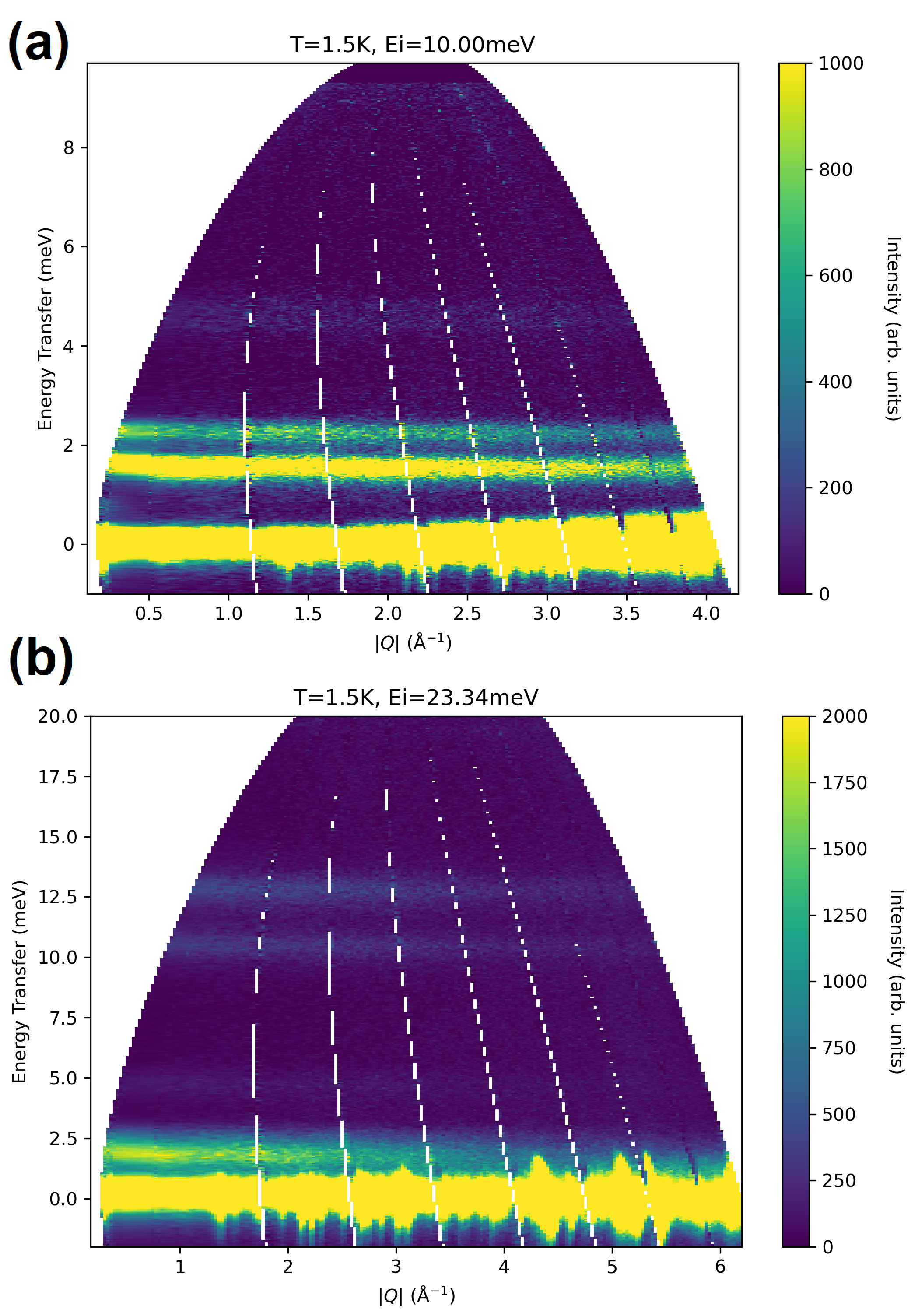}
\caption{Powder INS data (MARI, ISIS) for \ch{Ba3Tb(BO3)3} at $T=1.5$~K. (a) $E_\mathrm{i}=10$~meV, (b) $E_\mathrm{i}=23$~meV.}
\label{fig:insplot}
\end{figure}

\begin{figure}[htbp] 
\centering
\includegraphics[width=8cm]{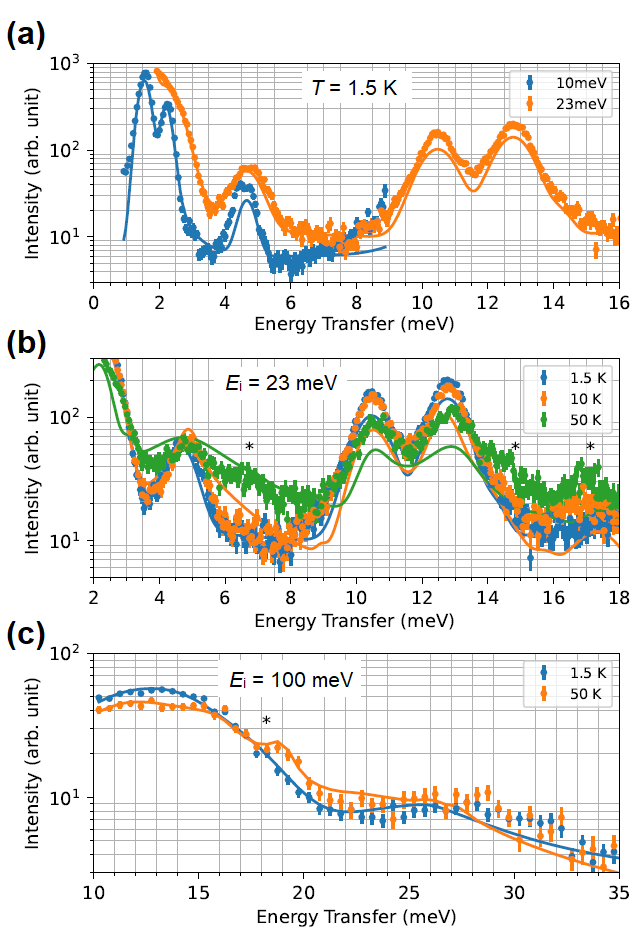}
\caption{Powder INS data for \ch{Ba3Tb(BO3)3}. Cuts along $\Delta E$ at 1.5~K (a), at different temperatures with (b) $E_\mathrm{i}$=23~meV and (c) $E_\mathrm{i}$=100~meV. Circles are measurements with vertical dashes indicating the $1\sigma$ uncertainty, solid lines are fits using the CEF model described in the text. The asterisks indicate energy transitions out of higher energy states than the ground state.}
\label{fig:inscuts}
\end{figure}

In order to determine the crystal electric field (CEF)
of the two Tb$^{3+}$ sites, INS data were collected at 1.5~K with $E_\mathrm{i}=10$ and 23~meV, Figures~\ref{fig:insplot} and~\ref{fig:inscuts}(a,b). The broad elastic channel has intensity independent of $Q$. Above this level, the first five CEF energy levels visible are 1.6, 2.3, 4.7, 10.5 and 12.9~meV, with intensities which decrease as a function of $Q$, as expected for magnetic bands. The data show no dispersion, which could mean that the Tb ions are too far apart (Tb--Tb distance $\approx5.45$~\AA) for spin correlations and the system is a simple paramagnet. However, our previous magnetic susceptibility measurements \cite{Kelly2020b} gave a non-zero Curie-Weiss temperature of $-9.8$~K which suggests that there are antiferromagnetic spin-spin correlations present in \ch{Ba3Tb(BO3)3}. Higher-energy data shown in Fig.~\ref{fig:inscuts}(c) shows another broad, weak peak at around 26~meV.

Individual cuts of the data were taken along $\Delta E$ to examine the temperature dependence of the spectrum, Fig.~\ref{fig:inscuts}(b). Peaks corresponding to transitions from the ground state, as listed above, are seen to decrease in intensity as the temperature is raised. On the other hand, there are possible peaks centered at energy transfer $E=6.6$, 14.7 and 17-19~meV, marked with an asterisk, which are not visible at $T=1.5$~K but grow in intensity from 10--50~K. Therefore, these values represent energy transitions from higher energy states than the ground state. The transition at 17~meV represents a transition from the first excited state, 1.6--18.6~meV. Based on the temperatures at which the peaks first appear, the excitation at 6.6~meV represents a transition from the second excited state (at 2.3~meV) and the excitation at 14.7~meV represents a transition from the third excited state (at 4.7~meV). The 17-18~meV peak, which appears around 17~meV in the $E_i=23$~meV data and 18~meV in the $E_i=100$~meV data only appears at 50~K and is probably an excited state transition from the 4.7~meV level to the $\approx$26~meV level. A schematic illustration of the energy levels is provided in the Supplemental Material \cite{supplemental}.

The Tb$^{3+}$ ions have electronic configuration $4f^8$, $J=6$, so they are non-Kramers ions with an even number of electrons and integer spin, meaning that the surrounding crystal electric field can fully lift the degeneracy of the energy levels. The Tb$^{3+}$ ions occupy two distinct crystallographic sites in the material \cite{Kelly2020b}: one-third in site Tb1 (Wyckoff symbol $2a$, $\mathrm{C_{3v}}$ symmetry) and two-thirds in site Tb2 (Wyckoff symbol $4b$, C$_3$ symmetry) with different local arrangements of oxide ions, i.e., crystal electric field (CEF). These site symmetries are identical to those in \ch{TbInO3} \cite{Clark2019} which has the same crystallographic space group and the same arrangement of Tb$^{3+}$ ions as \ch{Ba3Tb(BO3)3}. With respect to the CEF splitting, $\mathrm{C_{3v}}$ and $\mathrm{C_3}$ symmetries are equivalent, so each of the $J=6$ levels on each site thus splits into 5 singlets and 4 doublets, although with different irreducible representations:
\begin{equation}
\mathrm{3 A_1 + 2 A_2 + 4 E\;(C_{3v})}
\end{equation}
\begin{equation}
\mathrm{5 A + 4 E\;(C_3)}
\end{equation}

\noindent where the A irreps are singlets and E irreps are doublets. However, the two different symmetries permit different non-zero Stevens parameters \cite{Stevens1952} $B_l^m$, with $\mathrm{C_3}$ allowing negative $m$ parameters (corresponding to ``sine" tesseral harmonics) in addition to positive $m$ parameters (corresponding to ``cosine" tesseral harmonics) whereas $\mathrm{C_{3v}}$ allows only positive $m$ parameters. Thus there are 6 Stevens parameters to fit for the $\mathrm{C_{3v}}$ site (with $l=2,4,6$ and $m=0,3,6$ subject to $m<2l$), and an additional 9 parameters for the $\mathrm{C_3}$ site, for a total 15 CEF parameters. The situation is therefore rather complex, and to progress the analysis we had to employ a two-step approach, where we first used a point charge model to reduce the number of free parameters to obtain starting $B_l^m$ parameters for a local minimization.

For the point charge model, we used the ionic positions refined from the diffraction data above and allowed the effective charges of each ion type to vary. We included neighbors up to 11~\AA\ in the point charge summation and performed a series of grid searches fitting only the 1.5~K data for efficiency. We found that the effective charges on the oxygen ions are most important, and that in order to obtain a good description of the data we had to consider three different types of oxygen effective charges: $q_{O1}$ on the oxygen octahedral cage of the Tb1 site ($\mathrm{C_{3v}}$, sites O1 and O6), $q_{O2}$ on the oxygen octahedral cage of the Tb2 site ($\mathrm{C_3}$, sites O2 and O5), and $q_{Oi}$ on the intermediate barium-rich layer (sites O3 and O4). The best fit was found with effective charges: $q_{O1}=-0.85|e|$, $q_{O2}=-0.6|e|$, $q_{Oi}=-0.65|e|$, $q_{Tb}=+2.1|e|$, $q_{Ba}=+1.6|e|$, and $q_{B}=+0.3|e|$\,\footnote{While these effective charges are significantly different from the formal charges on each ion, this is not unexpected given the significant covalency in the system, i.e.~the presence of \ch{BO3} molecular anions. In particular, one would expect the effective charge on the B atoms to be much lower than that of the more ionic Tb atoms, which is found to be the case.}, which yields a singlet ground state and first excited doublets on both sites with energies around 2~meV. Both sites also have excited doublets around 5~meV and the Tb1 site has an excited singlet at 10~meV and the Tb2 site an excited singlet at 12~meV in agreement with the data. 

\begin{table}[htbp]
\caption{CEF parameters $B^l_m = A^l_m \langle r^l \rangle \theta_l$ in Stevens normalization \cite{Stevens1952} in meV using the ``neutron scattering" convention where the Stevens operator equivalent factor $\theta_l$ and radial expectation value $\langle r^l \rangle$ are folded into the parameter. Negative $m$ parameters are coefficients of the ``sine" tesseral harmonics.}
\label{table:cefpars}
\begin{ruledtabular}
\begin{tabular}{c c c}
           & Tb1 ($C_{3v}$) (meV) & Tb2 ($C_3$) (meV) \\ \hline
$B^2_0$    &  0.005269            &  0.1385           \\
$B^4_0$    & -0.001262            &  0.001059         \\
$B^4_3$    &  0.02957             &  0.006059         \\
$B^6_0$    & -3.811e-6            & -1.348e-7         \\
$B^6_3$    & -4.196e-5            &  4.129e-5         \\
$B^6_6$    & -2.385e-4            &  2.250e-6         \\
$B^4_{-3}$ &                      & -0.01129          \\
$B^6_{-3}$ &                      & -3.576e-5         \\
$B^6_{-6}$ &                      & -1.983e-4
\end{tabular}
\end{ruledtabular}
\end{table}

Starting from this fit to the point charge model, all 15 Stevens parameters were refined on the full dataset at all temperatures using the Powell method as implemented in the {\tt scipy.optimize} Python package~\cite{scipy}, giving the parameters shown in Table~\ref{table:cefpars} and the solid lines in Figure~\ref{fig:inscuts}. Whilst the fit agrees well with the measurements at 1.5~K, it is less good at higher temperatures. This may be because we used starting parameters based on fitting just the 1.5~K data and may therefore have been locked into a local minimum. Alternatively, there may have been some thermalization issue with the sample such that its true temperature was lower than recorded by the sensor. If this were the case, the transitions from the ground state would be more intense than calculated (which can be seen in Fig.~\ref{fig:inscuts}) whilst the excited state transitions would be lower than calculated (only the case for the 6.6~meV peak). 

Based on the fitted parameters, the ground state of \ch{Ba3Tb(BO3)3} is a singlet on both the Tb1 and Tb2 sites with an excited doublet at 2.2~meV (1.5~meV) on the Tb1 (Tb2) site. On the Tb1 site the ground state wavefunction is approximately $0.66|J_z=0\rangle + 0.51|J_z=\pm3\rangle$, whilst on the Tb2 site it is $\approx 0.7|J_z=\pm3\rangle + 0.2|J_z=0\rangle$, indicating a weak planar-type single-ion anisotropy, with the $c$ axis being the magnetically hard axis. In contrast, although the isostructural compound \ch{Ba3Er(BO3)3} also displayed $XY$ spin anisotropy, its INS data indicated a ground state Kramers doublet and low-lying first excited doublet \cite{Ennis2024}. To our knowledge, no INS measurements have yet been carried out on \ch{Ba3Yb(BO3)3}, but experimental evidence points to a Kramers doublet ground state for \ch{Ba3Yb(BO3)3} \cite{Bag2021} and a crystal field gap to the first excited state of $\Delta \approx17$~meV on one Yb site and 107~meV on the other \cite{Jiang2022}. These differences in CEF levels between the different isostructural \ch{Ba3\textit{Ln}(BO3)3} compounds are likely to be the driving force for the significant differences in heat capacity data (see following section~\ref{section:hc}).

\subsection{Specific heat} \label{section:hc}
\begin{figure}[htbp] 
\centering
\includegraphics[width=8cm]{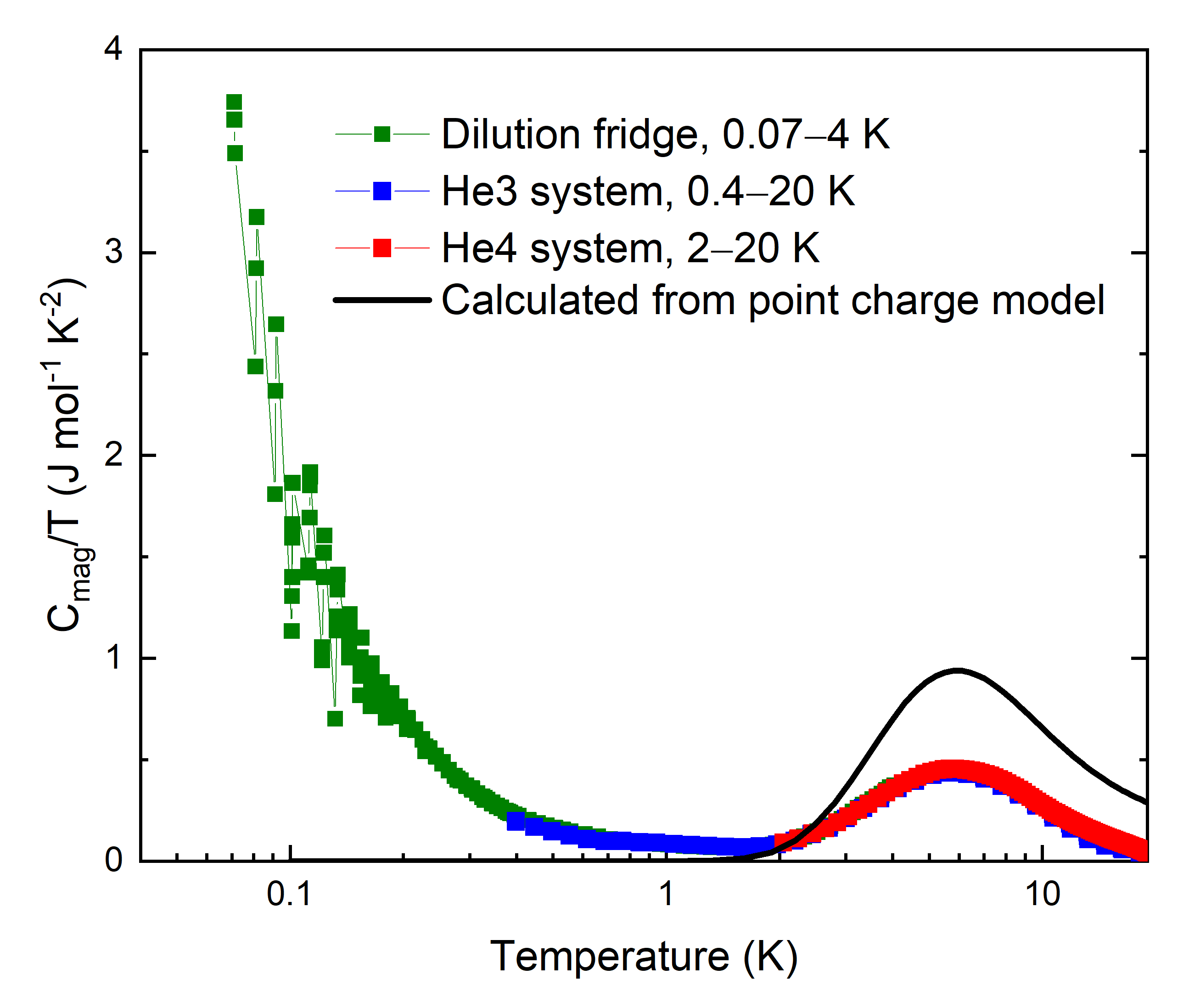}
\caption{Experimental heat capacity $C_\mathrm{p}/T$ of \ch{Ba3Tb(BO3)3} ($P6_3cm$) after subtraction of the Ag and \ch{Ba3Tb(BO3)3} lattice contributions using a Debye model. The black line shows the calculated magnetic heat capacity using the point charge model described in section~\ref{section:ins}.}
\label{fig:batbhc}
\end{figure}

Fig.~\ref{fig:batbhc} shows the zero-field heat capacity of \ch{Ba3Tb(BO3)3} as a function of temperature, after subtraction of the lattice contribution. The data at 2--20~K were reported previously \cite{Kelly2020b}. The measurements from three separate runs, on different portions of the same pellet, show excellent agreement. The magnetic heat capacity displays a broad peak at $T\approx 6$~K which is consistent with a hypothesis of short-range ordering in this temperature range \cite{Kelly2020b}. The magnetic heat capacity was also calculated from the point charge model discussed above (section~\ref{section:ins}). It is shown as a black line in Fig.~\ref{fig:batbhc} and its shape agrees very well with the experimental data. The discrepancy in magnitude can probably be explained by the fact that, in order to adequately fit the INS data, the point charge model needed to use charges that were different from the formal charges on the ions. However, our new INS data suggest that this peak is likely to result from thermal population of the lowest-lying CEF level at 1.6~meV. A similar peak centered at $T=6.5$~K was observed in the specific heat data of Ennis \textit{et al}.~for \ch{Ba3Er(BO3)3} single crystals \cite{Ennis2024}. The previously reported inverse magnetocaloric effect of \ch{Ba3Tb(BO3)3} at low temperatures \cite{Kelly2020b} might also be attributed to crystal field effects as it occurs on a similar energy scale.

The upturn in specific heat below 1~K may be attributed to the ordering of nuclear magnetic moments through the hyperfine interaction, i.e., the nuclear Schottky anomaly \cite{Hill1986,Mukherjee2017a,Hammann1973,Mirebeau2005} or could indicate the onset of a magnetic ordering transition below the lowest measured temperature of $\approx 70$~mK. The former explanation seems more probable given the absence of magnetic diffraction intensity observed by PND in the same temperature range. Future work could confirm this conclusion through collection of additional heat capacity data in applied magnetic fields. This would be expected to broaden the peak and shift it to higher temperatures; the data might then be fitted to a two-level Schottky function in order to extract values of the Zeeman splitting and saturation magnetic moment \cite{Pan2021}.

The isostructural \ch{Ba3Er(BO3)3} showed a sharp peak in $C_{mag}$ typical of antiferromagnetic long-range ordering at around 100~mK. Such a feature is completely absent in our study of \ch{Ba3Tb(BO3)3} and in Bag~\textit{et al}.'s study of \ch{Ba3Yb(BO3)3}, which they concluded was evidence for the absence of long-range order down to 56~mK \cite{Bag2021}. Jiang \textit{et al}.~also studied \ch{Ba3Yb(BO3)3} and argue that the increase of $C_{mag}/T$ below 1~K is due to electronic spin excitations \cite{Jiang2022}. Whilst we also believe that the Tb compound does not display LRO, we attribute the broad peak at higher temperatures in our sample (absent in the zero-field data for \ch{Ba3Yb(BO3)3}) to CEF effects which, as discussed above, have now been shown to differ dramatically between these three isostructural systems.

\subsection{Muon-spin relaxation spectroscopy}
\begin{figure*}[htbp]
\centering
\includegraphics[width=0.9\textwidth]{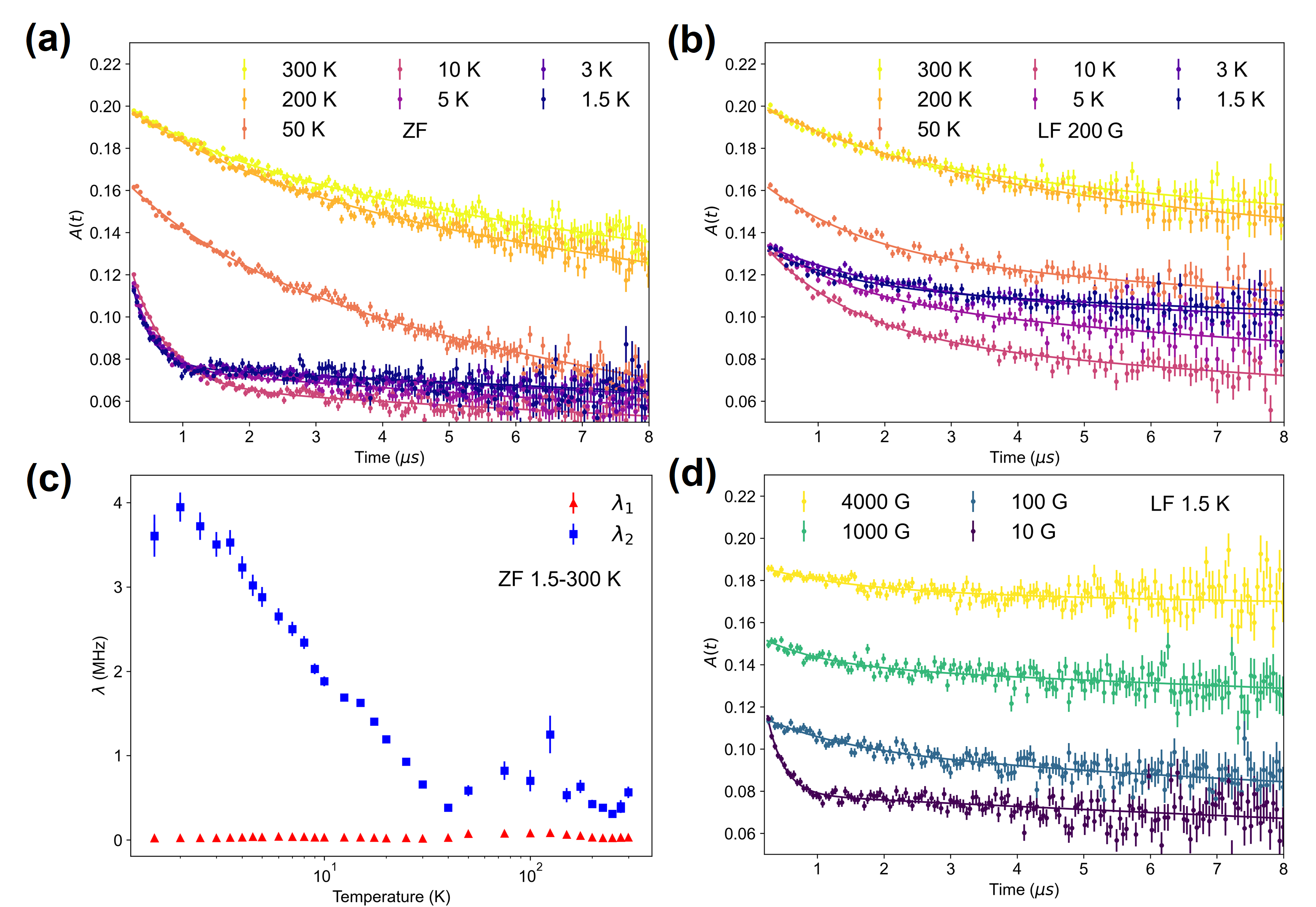}
\caption{(a) Fitted ZF $\mu$SR data at different temperatures. b) Fitted LF $\mu$SR data measured at 200~G. (c) Trends in the two relaxation rates $\lambda_1$ and $\lambda_2$ with temperature. (d) LF spectra in fields 10--4000~G at the lowest temperature studied ($1.5$~K).}
\label{fig:mu-final}
\end{figure*}

Muon-spin relaxation ($\mu$SR) data were measured on \ch{Ba3Tb(BO3)3} in zero field (ZF) and longitudinal field (LF) at several temperatures. The ZF data at selected temperatures are presented in Fig.~\ref{fig:mu-final}(a). The data are well described at all temperatures by equation \ref{eq:mu-ZF}:

\begin{equation}
A(t) = A_1 e^{ - \lambda_1t} + A_2 e^{ -\lambda_2t} + C
\label{eq:mu-ZF}
\end{equation}

\noindent where $C$ is a constant background term found to be 0.04, $A_1$ and $A_2$ are the component amplitudes and $\lambda_1$ and $\lambda_2$ the relaxation rates of the two exponential decay terms. The two relaxation rates of the fit can be seen in Fig.~\ref{fig:mu-final}(c), and can be clearly distinguished into a faster relaxing component $\lambda_1$ which varies with temperature and a slower background with a mean value of $\bar{\lambda}_2 \approx 0.036(3) $, which is attributed to muons stopping in the silver envelope.

The ZF asymmetry oscillations which are expected for a magnetically ordered system, are absent in data seen in Fig.~\ref{fig:mu-final}(a), confirming the absence of long-range magnetic order at all temperatures studied with $\mu$SR. The exponential decay form of the fitting equation is an indication that muon spin relaxation is primarily caused by rapid dynamic electronic spin fluctuations \cite{Mustonen2018}, suggesting a dynamic ground state. The further lack of a residual ``1/3" polarization tail in the asymmetry at long times allows us to exclude the presence of a magnetically frozen static spin system \cite{Khatua2022}.

Analysis of LF spectra in Fig.~\ref{fig:mu-final}(b) shows further proof of rapid electronic spin dynamics dominating the muon relaxation, as well as corroborating the lack of long-range ordering. Longitudinal fields can decouple and suppress relaxation caused by static nuclear fields from the larger electronic dynamic fields \cite{Nuccio2014}. The lack of significant changes in the ZF and LF asymmetry spectra in Fig.~\ref{fig:mu-final}(a) and (b), as well as the increase of the depolarization with decreasing LF field, as illustrated in Fig.~\ref{fig:mu-final}(d), are typical of a muon decay regime caused by fast fluctuating local electronic fields. Thus the analysis of $\mu$SR data collected on \ch{Ba3Tb(BO3)3} confirms the absence of LRO in the temperature range $1.5-300$ K as observed by magnetic susceptibility and heat capacity \cite{Kelly2020b}. 

Previous $\mu$SR data from a single crystal of the isostructural Ising system, \ch{Ba3Yb(BO3)3}, also showed an absence of oscillations, initial asymmetry loss, or a 1/3 polarization tail, indicating that the system shows no LRO or spin freezing \cite{Jiang2022}. Their data also suggest dynamic magnetic fluctuations at low temperatures. However, while their $\mu$SR data were also modelled using a two-exponential decay, the authors found that the two relaxation rates $\lambda_1$ and $\lambda_2$ plateaued below $T\approx 30$~K and they assigned these to two distinct muon stopping sites in the material, related to the different \textit{Ln} sites, whereas we find a single relaxation rate from \ch{Ba3Tb(BO3)3} which increases steadily upon cooling to 1.5~K. These differences might be explained by the different ground states of the materials, i.e.~a singlet state for \ch{Ba3Tb(BO3)3} but a Kramers doublet for \ch{Ba3Yb(BO3)3}.

\section{Conclusions}
In conclusion, we have carried out specific heat, neutron diffraction, INS, and $\mu$SR measurements on powder samples of \ch{Ba3Tb(BO3)3} in its hexagonal polymorph. In the neutron diffraction we observe no magnetic Bragg peaks or diffuse magnetic scattering down to a nominal base temperature of 75~mK. The $\mu$SR data down to 1.5~K show no oscillations, which if present would be indicative of long-range magnetic ordering, nor any conclusive evidence for spin freezing. The specific heat exhibits a broad hump at $T\approx 6$~K and an upturn towards the lowest temperatures measured, but no sharp $\lambda$-type peak or divergence. INS data provide evidence that this broad peak comes from thermal population of low-lying excited CEF levels, and fitting of the spectrum suggests a ground state singlet on both Tb sites with a doublet first excited state. We consider \ch{Ba3Tb(BO3)3} a worthwhile compound for further study in order to enable comparisons between this non-Kramers system and its isostructural Kramers analogues \ch{Ba3Yb(BO3)3} and \ch{Ba3Er(BO3)3}. Our future efforts will focus on the growth of single crystals in order to make a detailed study of the magnetic anisotropy.

\begin{acknowledgments}
This work is partly based on experiments performed at the Swiss spallation neutron source SINQ, Paul Scherrer Institute, Villigen, Switzerland. We acknowledge funding from the EPSRC for PhD studentships (EP/R513180/1 and EP/T517847/1) and the use of the Advanced Materials Characterisation Suite (EP/M000524/1). N.D.K.~acknowledges funding from Jesus College, Cambridge for a Research Fellowship. C.~T.~acknowledges a scholarship from the Winton Programme for the Physics of Sustainability. S.E.D.~acknowledges funding from the EPSRC (EP/T028580/1). Experiments at the ISIS Neutron and Muon Source were supported by beamtime allocations RB2000176 and RB2000251 from the Science and Technology Facilities Council (STFC). Data are available online \cite{Kelly2020MARI,Kelly2021EMU}. We thank STFC for access to the ISIS Materials Characterisation Laboratory. Data related to this publication will be made available in the Cambridge University Repository.
\end{acknowledgments}


\begin{thebibliography}{45}%
\makeatletter
\providecommand \@ifxundefined [1]{%
 \@ifx{#1\undefined}
}%
\providecommand \@ifnum [1]{%
 \ifnum #1\expandafter \@firstoftwo
 \else \expandafter \@secondoftwo
 \fi
}%
\providecommand \@ifx [1]{%
 \ifx #1\expandafter \@firstoftwo
 \else \expandafter \@secondoftwo
 \fi
}%
\providecommand \natexlab [1]{#1}%
\providecommand \enquote  [1]{``#1''}%
\providecommand \bibnamefont  [1]{#1}%
\providecommand \bibfnamefont [1]{#1}%
\providecommand \citenamefont [1]{#1}%
\providecommand \href@noop [0]{\@secondoftwo}%
\providecommand \href [0]{\begingroup \@sanitize@url \@href}%
\providecommand \@href[1]{\@@startlink{#1}\@@href}%
\providecommand \@@href[1]{\endgroup#1\@@endlink}%
\providecommand \@sanitize@url [0]{\catcode `\\12\catcode `\$12\catcode `\&12\catcode `\#12\catcode `\^12\catcode `\_12\catcode `\%12\relax}%
\providecommand \@@startlink[1]{}%
\providecommand \@@endlink[0]{}%
\providecommand \url  [0]{\begingroup\@sanitize@url \@url }%
\providecommand \@url [1]{\endgroup\@href {#1}{\urlprefix }}%
\providecommand \urlprefix  [0]{URL }%
\providecommand \Eprint [0]{\href }%
\providecommand \doibase [0]{https://doi.org/}%
\providecommand \selectlanguage [0]{\@gobble}%
\providecommand \bibinfo  [0]{\@secondoftwo}%
\providecommand \bibfield  [0]{\@secondoftwo}%
\providecommand \translation [1]{[#1]}%
\providecommand \BibitemOpen [0]{}%
\providecommand \bibitemStop [0]{}%
\providecommand \bibitemNoStop [0]{.\EOS\space}%
\providecommand \EOS [0]{\spacefactor3000\relax}%
\providecommand \BibitemShut  [1]{\csname bibitem#1\endcsname}%
\let\auto@bib@innerbib\@empty
\bibitem [{\citenamefont {Vojta}(2018)}]{Vojta2018}%
  \BibitemOpen
  \bibfield  {author} {\bibinfo {author} {\bibfnamefont {M.}~\bibnamefont {Vojta}},\ }\bibfield  {title} {\bibinfo {title} {{Frustration and quantum criticality}},\ }\href {https://doi.org/10.1088/1361-6633/aab6be} {\bibfield  {journal} {\bibinfo  {journal} {Rep. Prog. Phys.}\ }\textbf {\bibinfo {volume} {81}},\ \bibinfo {pages} {064501} (\bibinfo {year} {2018})}\BibitemShut {NoStop}%
\bibitem [{\citenamefont {Shores}\ \emph {et~al.}(2005)\citenamefont {Shores}, \citenamefont {Nytko}, \citenamefont {Bartlett},\ and\ \citenamefont {Nocera}}]{Shores2005}%
  \BibitemOpen
  \bibfield  {author} {\bibinfo {author} {\bibfnamefont {M.~P.}\ \bibnamefont {Shores}}, \bibinfo {author} {\bibfnamefont {E.~A.}\ \bibnamefont {Nytko}}, \bibinfo {author} {\bibfnamefont {B.~M.}\ \bibnamefont {Bartlett}},\ and\ \bibinfo {author} {\bibfnamefont {D.~G.}\ \bibnamefont {Nocera}},\ }\bibfield  {title} {\bibinfo {title} {{A structurally perfect $S=\nicefrac{1}{2}$ Kagom\'{e} antiferromagnet}},\ }\href {https://doi.org/10.1021/ja053891p} {\bibfield  {journal} {\bibinfo  {journal} {J. Am. Chem. Soc.}\ }\textbf {\bibinfo {volume} {127}},\ \bibinfo {pages} {13462} (\bibinfo {year} {2005})}\BibitemShut {NoStop}%
\bibitem [{\citenamefont {Vasiliev}\ \emph {et~al.}(2018)\citenamefont {Vasiliev}, \citenamefont {Volkova}, \citenamefont {Zvereva},\ and\ \citenamefont {Markina}}]{Vasiliev2018}%
  \BibitemOpen
  \bibfield  {author} {\bibinfo {author} {\bibfnamefont {A.}~\bibnamefont {Vasiliev}}, \bibinfo {author} {\bibfnamefont {O.}~\bibnamefont {Volkova}}, \bibinfo {author} {\bibfnamefont {E.}~\bibnamefont {Zvereva}},\ and\ \bibinfo {author} {\bibfnamefont {M.}~\bibnamefont {Markina}},\ }\bibfield  {title} {\bibinfo {title} {{Milestones of low-D quantum magnetism}},\ }\href {https://doi.org/10.1038/s41535-018-0090-7} {\bibfield  {journal} {\bibinfo  {journal} {npj Quantum Mater.}\ }\textbf {\bibinfo {volume} {3}},\ \bibinfo {pages} {18} (\bibinfo {year} {2018})}\BibitemShut {NoStop}%
\bibitem [{\citenamefont {Li}\ \emph {et~al.}(2015)\citenamefont {Li}, \citenamefont {Liao}, \citenamefont {Zhang}, \citenamefont {Li}, \citenamefont {Jin}, \citenamefont {Ling}, \citenamefont {Zhang}, \citenamefont {Zou}, \citenamefont {Pi}, \citenamefont {Yang}, \citenamefont {Wang}, \citenamefont {Wu},\ and\ \citenamefont {Zhang}}]{Li2015}%
  \BibitemOpen
  \bibfield  {author} {\bibinfo {author} {\bibfnamefont {Y.}~\bibnamefont {Li}}, \bibinfo {author} {\bibfnamefont {H.}~\bibnamefont {Liao}}, \bibinfo {author} {\bibfnamefont {Z.}~\bibnamefont {Zhang}}, \bibinfo {author} {\bibfnamefont {S.}~\bibnamefont {Li}}, \bibinfo {author} {\bibfnamefont {F.}~\bibnamefont {Jin}}, \bibinfo {author} {\bibfnamefont {L.}~\bibnamefont {Ling}}, \bibinfo {author} {\bibfnamefont {L.}~\bibnamefont {Zhang}}, \bibinfo {author} {\bibfnamefont {Y.}~\bibnamefont {Zou}}, \bibinfo {author} {\bibfnamefont {L.}~\bibnamefont {Pi}}, \bibinfo {author} {\bibfnamefont {Z.}~\bibnamefont {Yang}}, \bibinfo {author} {\bibfnamefont {J.}~\bibnamefont {Wang}}, \bibinfo {author} {\bibfnamefont {Z.}~\bibnamefont {Wu}},\ and\ \bibinfo {author} {\bibfnamefont {Q.}~\bibnamefont {Zhang}},\ }\bibfield  {title} {\bibinfo {title} {{Gapless quantum spin liquid ground state in the two-dimensional spin-$\nicefrac{1}{2}$ triangular antiferromagnet \ch{YbMgGaO4}}},\ }\href {https://doi.org/10.1038/srep16419}
  {\bibfield  {journal} {\bibinfo  {journal} {Sci. Rep.}\ }\textbf {\bibinfo {volume} {5}},\ \bibinfo {pages} {16419} (\bibinfo {year} {2015})}\BibitemShut {NoStop}%
\bibitem [{\citenamefont {Xu}\ \emph {et~al.}(2016)\citenamefont {Xu}, \citenamefont {Zhang}, \citenamefont {Li}, \citenamefont {Yu}, \citenamefont {Hong}, \citenamefont {Zhang},\ and\ \citenamefont {Li}}]{Xu2016}%
  \BibitemOpen
  \bibfield  {author} {\bibinfo {author} {\bibfnamefont {Y.}~\bibnamefont {Xu}}, \bibinfo {author} {\bibfnamefont {J.}~\bibnamefont {Zhang}}, \bibinfo {author} {\bibfnamefont {Y.~S.}\ \bibnamefont {Li}}, \bibinfo {author} {\bibfnamefont {Y.~J.}\ \bibnamefont {Yu}}, \bibinfo {author} {\bibfnamefont {X.~C.}\ \bibnamefont {Hong}}, \bibinfo {author} {\bibfnamefont {Q.~M.}\ \bibnamefont {Zhang}},\ and\ \bibinfo {author} {\bibfnamefont {S.~Y.}\ \bibnamefont {Li}},\ }\bibfield  {title} {\bibinfo {title} {Absence of magnetic thermal conductivity in the quantum spin-liquid candidate \ch{YbMgGaO4}},\ }\href {https://doi.org/10.1103/PhysRevLett.117.267202} {\bibfield  {journal} {\bibinfo  {journal} {Phys. Rev. Lett.}\ }\textbf {\bibinfo {volume} {117}},\ \bibinfo {pages} {267202} (\bibinfo {year} {2016})}\BibitemShut {NoStop}%
\bibitem [{\citenamefont {Li}\ \emph {et~al.}(2016)\citenamefont {Li}, \citenamefont {Adroja}, \citenamefont {Biswas}, \citenamefont {Baker}, \citenamefont {Zhang}, \citenamefont {Liu}, \citenamefont {Tsirlin}, \citenamefont {Gegenwart},\ and\ \citenamefont {Zhang}}]{Li2016}%
  \BibitemOpen
  \bibfield  {author} {\bibinfo {author} {\bibfnamefont {Y.}~\bibnamefont {Li}}, \bibinfo {author} {\bibfnamefont {D.~T.}\ \bibnamefont {Adroja}}, \bibinfo {author} {\bibfnamefont {P.~K.}\ \bibnamefont {Biswas}}, \bibinfo {author} {\bibfnamefont {P.~J.}\ \bibnamefont {Baker}}, \bibinfo {author} {\bibfnamefont {Q.}~\bibnamefont {Zhang}}, \bibinfo {author} {\bibfnamefont {J.}~\bibnamefont {Liu}}, \bibinfo {author} {\bibfnamefont {A.~A.}\ \bibnamefont {Tsirlin}}, \bibinfo {author} {\bibfnamefont {P.}~\bibnamefont {Gegenwart}},\ and\ \bibinfo {author} {\bibfnamefont {Q.}~\bibnamefont {Zhang}},\ }\bibfield  {title} {\bibinfo {title} {{Muon Spin Relaxation Evidence for the U(1) Quantum Spin-Liquid Ground State in the Triangular Antiferromagnet \ch{YbMgGaO4}}},\ }\href {https://doi.org/10.1103/PhysRevLett.117.097201} {\bibfield  {journal} {\bibinfo  {journal} {Phys. Rev. Lett.}\ }\textbf {\bibinfo {volume} {117}},\ \bibinfo {pages} {097201} (\bibinfo {year} {2016})}\BibitemShut {NoStop}%
\bibitem [{\citenamefont {Paddison}\ \emph {et~al.}(2017)\citenamefont {Paddison}, \citenamefont {Daum}, \citenamefont {Dun}, \citenamefont {Ehlers}, \citenamefont {Liu}, \citenamefont {Stone}, \citenamefont {Zhou},\ and\ \citenamefont {Mourigal}}]{Paddison2017}%
  \BibitemOpen
  \bibfield  {author} {\bibinfo {author} {\bibfnamefont {J.~A.~M.}\ \bibnamefont {Paddison}}, \bibinfo {author} {\bibfnamefont {M.}~\bibnamefont {Daum}}, \bibinfo {author} {\bibfnamefont {Z.}~\bibnamefont {Dun}}, \bibinfo {author} {\bibfnamefont {G.}~\bibnamefont {Ehlers}}, \bibinfo {author} {\bibfnamefont {Y.}~\bibnamefont {Liu}}, \bibinfo {author} {\bibfnamefont {M.~B.}\ \bibnamefont {Stone}}, \bibinfo {author} {\bibfnamefont {H.}~\bibnamefont {Zhou}},\ and\ \bibinfo {author} {\bibfnamefont {M.}~\bibnamefont {Mourigal}},\ }\bibfield  {title} {\bibinfo {title} {{Continuous excitations of the triangular-lattice quantum spin liquid \ch{YbMgGaO4}}},\ }\href {https://doi.org/10.1038/nphys3971} {\bibfield  {journal} {\bibinfo  {journal} {Nat. Phys.}\ }\textbf {\bibinfo {volume} {13}},\ \bibinfo {pages} {117} (\bibinfo {year} {2017})}\BibitemShut {NoStop}%
\bibitem [{\citenamefont {Li}\ \emph {et~al.}(2017)\citenamefont {Li}, \citenamefont {Adroja}, \citenamefont {Bewley}, \citenamefont {Voneshen}, \citenamefont {Tsirlin}, \citenamefont {Gegenwart},\ and\ \citenamefont {Zhang}}]{Li2017}%
  \BibitemOpen
  \bibfield  {author} {\bibinfo {author} {\bibfnamefont {Y.}~\bibnamefont {Li}}, \bibinfo {author} {\bibfnamefont {D.~T.}\ \bibnamefont {Adroja}}, \bibinfo {author} {\bibfnamefont {R.~I.}\ \bibnamefont {Bewley}}, \bibinfo {author} {\bibfnamefont {D.}~\bibnamefont {Voneshen}}, \bibinfo {author} {\bibfnamefont {A.~A.}\ \bibnamefont {Tsirlin}}, \bibinfo {author} {\bibfnamefont {P.}~\bibnamefont {Gegenwart}},\ and\ \bibinfo {author} {\bibfnamefont {Q.}~\bibnamefont {Zhang}},\ }\bibfield  {title} {\bibinfo {title} {{Crystalline Electric-Field Randomness in the Triangular Lattice Spin-Liquid \ch{YbMgGaO4}}},\ }\href {https://doi.org/10.1103/PhysRevLett.118.107202} {\bibfield  {journal} {\bibinfo  {journal} {Phys. Rev. Lett.}\ }\textbf {\bibinfo {volume} {118}},\ \bibinfo {pages} {107202} (\bibinfo {year} {2017})}\BibitemShut {NoStop}%
\bibitem [{\citenamefont {Ma}\ \emph {et~al.}(2018)\citenamefont {Ma}, \citenamefont {Wang}, \citenamefont {Dong}, \citenamefont {Zhang}, \citenamefont {Li}, \citenamefont {Zheng}, \citenamefont {Yu}, \citenamefont {Wang}, \citenamefont {Che}, \citenamefont {Ran}, \citenamefont {Bao}, \citenamefont {Cai}, \citenamefont {{\v{C}}erm{\'{a}}k}, \citenamefont {Schneidewind}, \citenamefont {Yano}, \citenamefont {Gardner}, \citenamefont {Lu}, \citenamefont {Yu}, \citenamefont {Liu}, \citenamefont {Li}, \citenamefont {Li},\ and\ \citenamefont {Wen}}]{Ma2018}%
  \BibitemOpen
  \bibfield  {author} {\bibinfo {author} {\bibfnamefont {Z.}~\bibnamefont {Ma}}, \bibinfo {author} {\bibfnamefont {J.}~\bibnamefont {Wang}}, \bibinfo {author} {\bibfnamefont {Z.~Y.}\ \bibnamefont {Dong}}, \bibinfo {author} {\bibfnamefont {J.}~\bibnamefont {Zhang}}, \bibinfo {author} {\bibfnamefont {S.}~\bibnamefont {Li}}, \bibinfo {author} {\bibfnamefont {S.~H.}\ \bibnamefont {Zheng}}, \bibinfo {author} {\bibfnamefont {Y.}~\bibnamefont {Yu}}, \bibinfo {author} {\bibfnamefont {W.}~\bibnamefont {Wang}}, \bibinfo {author} {\bibfnamefont {L.}~\bibnamefont {Che}}, \bibinfo {author} {\bibfnamefont {K.}~\bibnamefont {Ran}}, \bibinfo {author} {\bibfnamefont {S.}~\bibnamefont {Bao}}, \bibinfo {author} {\bibfnamefont {Z.}~\bibnamefont {Cai}}, \bibinfo {author} {\bibfnamefont {P.}~\bibnamefont {{\v{C}}erm{\'{a}}k}}, \bibinfo {author} {\bibfnamefont {A.}~\bibnamefont {Schneidewind}}, \bibinfo {author} {\bibfnamefont {S.}~\bibnamefont {Yano}}, \bibinfo {author} {\bibfnamefont {J.~S.}\ \bibnamefont {Gardner}}, \bibinfo
  {author} {\bibfnamefont {X.}~\bibnamefont {Lu}}, \bibinfo {author} {\bibfnamefont {S.~L.}\ \bibnamefont {Yu}}, \bibinfo {author} {\bibfnamefont {J.~M.}\ \bibnamefont {Liu}}, \bibinfo {author} {\bibfnamefont {S.}~\bibnamefont {Li}}, \bibinfo {author} {\bibfnamefont {J.~X.}\ \bibnamefont {Li}},\ and\ \bibinfo {author} {\bibfnamefont {J.}~\bibnamefont {Wen}},\ }\bibfield  {title} {\bibinfo {title} {{Spin-Glass Ground State in a Triangular-Lattice Compound \ch{YbZnGaO4}}},\ }\href {https://doi.org/10.1103/PhysRevLett.120.087201} {\bibfield  {journal} {\bibinfo  {journal} {Phys. Rev. Lett.}\ }\textbf {\bibinfo {volume} {120}},\ \bibinfo {pages} {087201} (\bibinfo {year} {2018})}\BibitemShut {NoStop}%
\bibitem [{\citenamefont {Khamaganova}\ \emph {et~al.}(1999)\citenamefont {Khamaganova}, \citenamefont {Kuperman},\ and\ \citenamefont {Bazarova}}]{Khamaganova1999}%
  \BibitemOpen
  \bibfield  {author} {\bibinfo {author} {\bibfnamefont {T.~N.}\ \bibnamefont {Khamaganova}}, \bibinfo {author} {\bibfnamefont {N.~M.}\ \bibnamefont {Kuperman}},\ and\ \bibinfo {author} {\bibfnamefont {Z.~G.}\ \bibnamefont {Bazarova}},\ }\bibfield  {title} {\bibinfo {title} {{The Double Borates \ch{Ba3\textit{Ln}(BO3)3}, \textit{Ln}=La-Lu, Y}},\ }\href {https://doi.org/10.1006/jssc.1999.8163} {\bibfield  {journal} {\bibinfo  {journal} {J. Solid State Chem.}\ }\textbf {\bibinfo {volume} {145}},\ \bibinfo {pages} {33} (\bibinfo {year} {1999})}\BibitemShut {NoStop}%
\bibitem [{\citenamefont {Simura}\ \emph {et~al.}(2017)\citenamefont {Simura}, \citenamefont {Kawai},\ and\ \citenamefont {Sugiyama}}]{Simura2017}%
  \BibitemOpen
  \bibfield  {author} {\bibinfo {author} {\bibfnamefont {R.}~\bibnamefont {Simura}}, \bibinfo {author} {\bibfnamefont {S.}~\bibnamefont {Kawai}},\ and\ \bibinfo {author} {\bibfnamefont {K.}~\bibnamefont {Sugiyama}},\ }\bibfield  {title} {\bibinfo {title} {{Phase Transition and Thermal Expansion of \ch{Ba3\textit{R}B3O9} (\textit{R} = Sm-Yb, and Y)}},\ }\href {https://doi.org/10.1515/htmp-2015-0290} {\bibfield  {journal} {\bibinfo  {journal} {High Temp. Mater. Process.}\ }\textbf {\bibinfo {volume} {36}},\ \bibinfo {pages} {763} (\bibinfo {year} {2017})}\BibitemShut {NoStop}%
\bibitem [{\citenamefont {Gao}\ \emph {et~al.}(2018)\citenamefont {Gao}, \citenamefont {Xu}, \citenamefont {Tian},\ and\ \citenamefont {Yuan}}]{Gao2018}%
  \BibitemOpen
  \bibfield  {author} {\bibinfo {author} {\bibfnamefont {Y.}~\bibnamefont {Gao}}, \bibinfo {author} {\bibfnamefont {L.}~\bibnamefont {Xu}}, \bibinfo {author} {\bibfnamefont {Z.}~\bibnamefont {Tian}},\ and\ \bibinfo {author} {\bibfnamefont {S.}~\bibnamefont {Yuan}},\ }\bibfield  {title} {\bibinfo {title} {{Synthesis and magnetism of \ch{RE(BaBO3)3} (RE=Dy,Ho,Er,Tm,Yb) series with rare earth ions on a two dimensional triangle-lattice}},\ }\href {https://doi.org/10.1016/j.jallcom.2018.02.110} {\bibfield  {journal} {\bibinfo  {journal} {J. Alloys Compd.}\ }\textbf {\bibinfo {volume} {745}},\ \bibinfo {pages} {396} (\bibinfo {year} {2018})}\BibitemShut {NoStop}%
\bibitem [{\citenamefont {Zeng}\ \emph {et~al.}(2020)\citenamefont {Zeng}, \citenamefont {Ma}, \citenamefont {Gao}, \citenamefont {Tian}, \citenamefont {Ling},\ and\ \citenamefont {Pi}}]{Zeng2020}%
  \BibitemOpen
  \bibfield  {author} {\bibinfo {author} {\bibfnamefont {K.~Y.}\ \bibnamefont {Zeng}}, \bibinfo {author} {\bibfnamefont {L.}~\bibnamefont {Ma}}, \bibinfo {author} {\bibfnamefont {Y.~X.}\ \bibnamefont {Gao}}, \bibinfo {author} {\bibfnamefont {Z.~M.}\ \bibnamefont {Tian}}, \bibinfo {author} {\bibfnamefont {L.~S.}\ \bibnamefont {Ling}},\ and\ \bibinfo {author} {\bibfnamefont {L.}~\bibnamefont {Pi}},\ }\bibfield  {title} {\bibinfo {title} {{NMR evidence for gapless quantum spin liquid state in the ideal triangular-lattice compound \ch{Yb(BaBO3)3}}},\ }\href@noop {} {\bibfield  {journal} {\bibinfo  {journal} {Phys. Rev. B}\ }\textbf {\bibinfo {volume} {102}},\ \bibinfo {pages} {045149} (\bibinfo {year} {2020})}\BibitemShut {NoStop}%
\bibitem [{\citenamefont {Cho}\ \emph {et~al.}(2021)\citenamefont {Cho}, \citenamefont {Blundell}, \citenamefont {Shiroka}, \citenamefont {Macfarquharson}, \citenamefont {Prabhakaran},\ and\ \citenamefont {Coldea}}]{Cho2021}%
  \BibitemOpen
  \bibfield  {author} {\bibinfo {author} {\bibfnamefont {H.}~\bibnamefont {Cho}}, \bibinfo {author} {\bibfnamefont {S.~J.}\ \bibnamefont {Blundell}}, \bibinfo {author} {\bibfnamefont {T.}~\bibnamefont {Shiroka}}, \bibinfo {author} {\bibfnamefont {K.}~\bibnamefont {Macfarquharson}}, \bibinfo {author} {\bibfnamefont {D.}~\bibnamefont {Prabhakaran}},\ and\ \bibinfo {author} {\bibfnamefont {R.}~\bibnamefont {Coldea}},\ }\bibfield  {title} {\bibinfo {title} {{Studies on Novel Yb-based Candidate Triangular Quantum Antiferromagnets: \ch{Ba3YbB3O9} and \ch{Ba3YbB9O18}}},\ }\href@noop {} {\bibfield  {journal} {\bibinfo  {journal} {arxiv:2104.01005}\ } (\bibinfo {year} {2021})}\BibitemShut {NoStop}%
\bibitem [{\citenamefont {Bag}\ \emph {et~al.}(2021)\citenamefont {Bag}, \citenamefont {Ennis}, \citenamefont {Liu}, \citenamefont {Dissanayake}, \citenamefont {Shi}, \citenamefont {Liu}, \citenamefont {Balents},\ and\ \citenamefont {Haravifard}}]{Bag2021}%
  \BibitemOpen
  \bibfield  {author} {\bibinfo {author} {\bibfnamefont {R.}~\bibnamefont {Bag}}, \bibinfo {author} {\bibfnamefont {M.}~\bibnamefont {Ennis}}, \bibinfo {author} {\bibfnamefont {C.}~\bibnamefont {Liu}}, \bibinfo {author} {\bibfnamefont {S.~E.}\ \bibnamefont {Dissanayake}}, \bibinfo {author} {\bibfnamefont {Z.}~\bibnamefont {Shi}}, \bibinfo {author} {\bibfnamefont {J.}~\bibnamefont {Liu}}, \bibinfo {author} {\bibfnamefont {L.}~\bibnamefont {Balents}},\ and\ \bibinfo {author} {\bibfnamefont {S.}~\bibnamefont {Haravifard}},\ }\bibfield  {title} {\bibinfo {title} {{Realization of Quantum Dipoles in Triangular Lattice Crystal \ch{Ba3Yb(BO3)3}}},\ }\href@noop {} {\bibfield  {journal} {\bibinfo  {journal} {Phys. Rev. B}\ }\textbf {\bibinfo {volume} {104}},\ \bibinfo {pages} {L220403} (\bibinfo {year} {2021})}\BibitemShut {NoStop}%
\bibitem [{\citenamefont {Ennis}\ \emph {et~al.}(2024)\citenamefont {Ennis}, \citenamefont {Bag}, \citenamefont {Liu}, \citenamefont {Dissanayake}, \citenamefont {Kolesnikov}, \citenamefont {Balents},\ and\ \citenamefont {Haravifard}}]{Ennis2024}%
  \BibitemOpen
  \bibfield  {author} {\bibinfo {author} {\bibfnamefont {M.}~\bibnamefont {Ennis}}, \bibinfo {author} {\bibfnamefont {R.}~\bibnamefont {Bag}}, \bibinfo {author} {\bibfnamefont {C.}~\bibnamefont {Liu}}, \bibinfo {author} {\bibfnamefont {S.~E.}\ \bibnamefont {Dissanayake}}, \bibinfo {author} {\bibfnamefont {A.~I.}\ \bibnamefont {Kolesnikov}}, \bibinfo {author} {\bibfnamefont {L.}~\bibnamefont {Balents}},\ and\ \bibinfo {author} {\bibfnamefont {S.}~\bibnamefont {Haravifard}},\ }\bibfield  {title} {\bibinfo {title} {{Realization of two-sublattice exchange physics in the triangular lattice compound \ch{Ba3Er(BO3)3}}},\ }\href {https://doi.org/10.1038/s42005-024-01532-w} {\bibfield  {journal} {\bibinfo  {journal} {Commun. Phys.}\ }\textbf {\bibinfo {volume} {7}},\ \bibinfo {pages} {37} (\bibinfo {year} {2024})}\BibitemShut {NoStop}%
\bibitem [{\citenamefont {Momma}\ and\ \citenamefont {Izumi}(2011)}]{Momma2011}%
  \BibitemOpen
  \bibfield  {author} {\bibinfo {author} {\bibfnamefont {K.}~\bibnamefont {Momma}}\ and\ \bibinfo {author} {\bibfnamefont {F.}~\bibnamefont {Izumi}},\ }\bibfield  {title} {\bibinfo {title} {{VESTA 3 for three-dimensional visualization of crystal, volumetric and morphology data}},\ }\href {https://doi.org/10.1107/S0021889811038970} {\bibfield  {journal} {\bibinfo  {journal} {J. Appl. Crystallogr.}\ }\textbf {\bibinfo {volume} {44}},\ \bibinfo {pages} {1272} (\bibinfo {year} {2011})}\BibitemShut {NoStop}%
\bibitem [{\citenamefont {Kelly}\ \emph {et~al.}(2020{\natexlab{a}})\citenamefont {Kelly}, \citenamefont {Liu},\ and\ \citenamefont {Dutton}}]{Kelly2020b}%
  \BibitemOpen
  \bibfield  {author} {\bibinfo {author} {\bibfnamefont {N.~D.}\ \bibnamefont {Kelly}}, \bibinfo {author} {\bibfnamefont {C.}~\bibnamefont {Liu}},\ and\ \bibinfo {author} {\bibfnamefont {S.~E.}\ \bibnamefont {Dutton}},\ }\bibfield  {title} {\bibinfo {title} {{Structure and magnetism of a new hexagonal polymorph of \ch{Ba3Tb(BO3)3} with a quasi-2D triangular lattice}},\ }\href {https://doi.org/10.1016/j.jssc.2020.121640} {\bibfield  {journal} {\bibinfo  {journal} {J. Solid State Chem.}\ }\textbf {\bibinfo {volume} {292}},\ \bibinfo {pages} {121640} (\bibinfo {year} {2020}{\natexlab{a}})}\BibitemShut {NoStop}%
\bibitem [{\citenamefont {Ramirez}(1994)}]{Ramirez1994}%
  \BibitemOpen
  \bibfield  {author} {\bibinfo {author} {\bibfnamefont {A.~P.}\ \bibnamefont {Ramirez}},\ }\bibfield  {title} {\bibinfo {title} {{Strongly geometrically frustrated magnets}},\ }\href@noop {} {\bibfield  {journal} {\bibinfo  {journal} {Annu. Rev. Mater. Sci.}\ }\textbf {\bibinfo {volume} {24}},\ \bibinfo {pages} {453} (\bibinfo {year} {1994})}\BibitemShut {NoStop}%
\bibitem [{\citenamefont {Fischer}\ \emph {et~al.}(2000)\citenamefont {Fischer}, \citenamefont {Frey}, \citenamefont {Koch}, \citenamefont {K{\"o}nnecke}, \citenamefont {Pomjakushin}, \citenamefont {Schefer}, \citenamefont {Thut}, \citenamefont {Schlumpf}, \citenamefont {B{\"u}rge}, \citenamefont {Greuter}, \citenamefont {Bondt},\ and\ \citenamefont {Berruyer}}]{Fischer2000}%
  \BibitemOpen
  \bibfield  {author} {\bibinfo {author} {\bibfnamefont {P.}~\bibnamefont {Fischer}}, \bibinfo {author} {\bibfnamefont {G.}~\bibnamefont {Frey}}, \bibinfo {author} {\bibfnamefont {M.}~\bibnamefont {Koch}}, \bibinfo {author} {\bibfnamefont {M.}~\bibnamefont {K{\"o}nnecke}}, \bibinfo {author} {\bibfnamefont {V.}~\bibnamefont {Pomjakushin}}, \bibinfo {author} {\bibfnamefont {J.}~\bibnamefont {Schefer}}, \bibinfo {author} {\bibfnamefont {R.}~\bibnamefont {Thut}}, \bibinfo {author} {\bibfnamefont {N.}~\bibnamefont {Schlumpf}}, \bibinfo {author} {\bibfnamefont {R.}~\bibnamefont {B{\"u}rge}}, \bibinfo {author} {\bibfnamefont {U.}~\bibnamefont {Greuter}}, \bibinfo {author} {\bibfnamefont {S.}~\bibnamefont {Bondt}},\ and\ \bibinfo {author} {\bibfnamefont {E.}~\bibnamefont {Berruyer}},\ }\bibfield  {title} {\bibinfo {title} {High-resolution powder diffractometer HRPT for thermal neutrons at SINQ},\ }\href@noop {} {\bibfield  {journal} {\bibinfo  {journal} {Physica B}\ }\textbf {\bibinfo {volume} {276--278}},\ \bibinfo
  {pages} {146} (\bibinfo {year} {2000})}\BibitemShut {NoStop}%
\bibitem [{\citenamefont {Rietveld}(1969)}]{Rietveld1969}%
  \BibitemOpen
  \bibfield  {author} {\bibinfo {author} {\bibfnamefont {H.~M.}\ \bibnamefont {Rietveld}},\ }\bibfield  {title} {\bibinfo {title} {{A profile refinement method for nuclear and magnetic structures}},\ }\href {https://doi.org/10.1107/S0021889869006558} {\bibfield  {journal} {\bibinfo  {journal} {J. Appl. Crystallogr.}\ }\textbf {\bibinfo {volume} {2}},\ \bibinfo {pages} {65} (\bibinfo {year} {1969})}\BibitemShut {NoStop}%
\bibitem [{\citenamefont {Coelho}(2018)}]{Coelho2018}%
  \BibitemOpen
  \bibfield  {author} {\bibinfo {author} {\bibfnamefont {A.~A.}\ \bibnamefont {Coelho}},\ }\bibfield  {title} {\bibinfo {title} {{TOPAS and TOPAS-Academic: An optimization program integrating computer algebra and crystallographic objects written in C++}},\ }\href {https://doi.org/10.1107/S1600576718000183} {\bibfield  {journal} {\bibinfo  {journal} {J. Appl. Crystallogr.}\ }\textbf {\bibinfo {volume} {51}},\ \bibinfo {pages} {210} (\bibinfo {year} {2018})}\BibitemShut {NoStop}%
\bibitem [{\citenamefont {Young}(1993)}]{Young1993}%
  \BibitemOpen
  \bibinfo {editor} {\bibfnamefont {R.~A.}\ \bibnamefont {Young}},\ ed.,\ \href@noop {} {\emph {\bibinfo {title} {{The Rietveld Method}}}}\ (\bibinfo  {publisher} {Oxford University Press},\ \bibinfo {year} {1993})\BibitemShut {NoStop}%
\bibitem [{\citenamefont {Kelly}\ \emph {et~al.}(2020{\natexlab{b}})\citenamefont {Kelly}, \citenamefont {Dutton},\ and\ \citenamefont {Le}}]{Kelly2020MARI}%
  \BibitemOpen
  \bibfield  {author} {\bibinfo {author} {\bibfnamefont {N.~D.}\ \bibnamefont {Kelly}}, \bibinfo {author} {\bibfnamefont {S.~E.}\ \bibnamefont {Dutton}},\ and\ \bibinfo {author} {\bibfnamefont {M.~D.}\ \bibnamefont {Le}},\ }\bibfield  {title} {\bibinfo {title} {{Inelastic neutron spectrum of hexagonal \ch{Ba3Tb(BO3)3}}},\ }\bibfield  {journal} {\bibinfo  {journal} {STFC ISIS Neutron Muon Source}\ }\href {https://doi.org/10.5286/ISIS.E.RB2000176} {10.5286/ISIS.E.RB2000176} (\bibinfo {year} {2020}{\natexlab{b}})\BibitemShut {NoStop}%
\bibitem [{\citenamefont {Arblaster}(2015)}]{Arblaster2015}%
  \BibitemOpen
  \bibfield  {author} {\bibinfo {author} {\bibfnamefont {J.~W.}\ \bibnamefont {Arblaster}},\ }\bibfield  {title} {\bibinfo {title} {{Thermodynamic Properties of Silver}},\ }\href {https://doi.org/10.1007/s11669-015-0411-5} {\bibfield  {journal} {\bibinfo  {journal} {J. Phase Equilibria Diffus.}\ }\textbf {\bibinfo {volume} {36}},\ \bibinfo {pages} {573} (\bibinfo {year} {2015})}\BibitemShut {NoStop}%
\bibitem [{\citenamefont {Gopal}(1966)}]{Gopal1966}%
  \BibitemOpen
  \bibfield  {author} {\bibinfo {author} {\bibfnamefont {E.~S.~R.}\ \bibnamefont {Gopal}},\ }\href@noop {} {\emph {\bibinfo {title} {Specific Heats at Low Temperatures}}}\ (\bibinfo  {publisher} {Springer US},\ \bibinfo {address} {Boston, MA},\ \bibinfo {year} {1966})\BibitemShut {NoStop}%
\bibitem [{\citenamefont {Kelly}\ \emph {et~al.}(2021)\citenamefont {Kelly}, \citenamefont {Dutton},\ and\ \citenamefont {Baker}}]{Kelly2021EMU}%
  \BibitemOpen
  \bibfield  {author} {\bibinfo {author} {\bibfnamefont {N.~D.}\ \bibnamefont {Kelly}}, \bibinfo {author} {\bibfnamefont {S.~E.}\ \bibnamefont {Dutton}},\ and\ \bibinfo {author} {\bibfnamefont {P.~J.}\ \bibnamefont {Baker}},\ }\bibfield  {title} {\bibinfo {title} {{Muon spectroscopy to probe spin dynamics in hexagonal \ch{Ba3Tb(BO3)3}}},\ }\bibfield  {journal} {\bibinfo  {journal} {STFC ISIS Neutron Muon Source}\ }\href {https://doi.org/10.5286/ISIS.E.RB2000251} {10.5286/ISIS.E.RB2000251} (\bibinfo {year} {2021})\BibitemShut {NoStop}%
\bibitem [{\citenamefont {Arnold}\ \emph {et~al.}(2014)\citenamefont {Arnold}, \citenamefont {Bilheux}, \citenamefont {Borreguero}, \citenamefont {Buts}, \citenamefont {Campbell}, \citenamefont {Chapon}, \citenamefont {Doucet}, \citenamefont {Draper}, \citenamefont {{Ferraz Leal}}, \citenamefont {Gigg}, \citenamefont {Lynch}, \citenamefont {Markvardsen}, \citenamefont {Mikkelson}, \citenamefont {Mikkelson}, \citenamefont {Miller}, \citenamefont {Palmen}, \citenamefont {Parker}, \citenamefont {Passos}, \citenamefont {Perring}, \citenamefont {Peterson}, \citenamefont {Ren}, \citenamefont {Reuter}, \citenamefont {Savici}, \citenamefont {Taylor}, \citenamefont {Taylor}, \citenamefont {Tolchenov}, \citenamefont {Zhou},\ and\ \citenamefont {Zikovsky}}]{Arnold2014}%
  \BibitemOpen
  \bibfield  {author} {\bibinfo {author} {\bibfnamefont {O.}~\bibnamefont {Arnold}}, \bibinfo {author} {\bibfnamefont {J.~C.}\ \bibnamefont {Bilheux}}, \bibinfo {author} {\bibfnamefont {J.~M.}\ \bibnamefont {Borreguero}}, \bibinfo {author} {\bibfnamefont {A.}~\bibnamefont {Buts}}, \bibinfo {author} {\bibfnamefont {S.~I.}\ \bibnamefont {Campbell}}, \bibinfo {author} {\bibfnamefont {L.}~\bibnamefont {Chapon}}, \bibinfo {author} {\bibfnamefont {M.}~\bibnamefont {Doucet}}, \bibinfo {author} {\bibfnamefont {N.}~\bibnamefont {Draper}}, \bibinfo {author} {\bibfnamefont {R.}~\bibnamefont {{Ferraz Leal}}}, \bibinfo {author} {\bibfnamefont {M.~A.}\ \bibnamefont {Gigg}}, \bibinfo {author} {\bibfnamefont {V.~E.}\ \bibnamefont {Lynch}}, \bibinfo {author} {\bibfnamefont {A.}~\bibnamefont {Markvardsen}}, \bibinfo {author} {\bibfnamefont {D.~J.}\ \bibnamefont {Mikkelson}}, \bibinfo {author} {\bibfnamefont {R.~L.}\ \bibnamefont {Mikkelson}}, \bibinfo {author} {\bibfnamefont {R.}~\bibnamefont {Miller}}, \bibinfo {author}
  {\bibfnamefont {K.}~\bibnamefont {Palmen}}, \bibinfo {author} {\bibfnamefont {P.}~\bibnamefont {Parker}}, \bibinfo {author} {\bibfnamefont {G.}~\bibnamefont {Passos}}, \bibinfo {author} {\bibfnamefont {T.~G.}\ \bibnamefont {Perring}}, \bibinfo {author} {\bibfnamefont {P.~F.}\ \bibnamefont {Peterson}}, \bibinfo {author} {\bibfnamefont {S.}~\bibnamefont {Ren}}, \bibinfo {author} {\bibfnamefont {M.~A.}\ \bibnamefont {Reuter}}, \bibinfo {author} {\bibfnamefont {A.~T.}\ \bibnamefont {Savici}}, \bibinfo {author} {\bibfnamefont {J.~W.}\ \bibnamefont {Taylor}}, \bibinfo {author} {\bibfnamefont {R.~J.}\ \bibnamefont {Taylor}}, \bibinfo {author} {\bibfnamefont {R.}~\bibnamefont {Tolchenov}}, \bibinfo {author} {\bibfnamefont {W.}~\bibnamefont {Zhou}},\ and\ \bibinfo {author} {\bibfnamefont {J.}~\bibnamefont {Zikovsky}},\ }\bibfield  {title} {\bibinfo {title} {{Mantid - Data analysis and visualization package for neutron scattering and $\mu$ SR experiments}},\ }\href {https://doi.org/10.1016/j.nima.2014.07.029}
  {\bibfield  {journal} {\bibinfo  {journal} {Nucl. Instruments Methods Phys. Res. Sect. A Accel. Spectrometers, Detect. Assoc. Equip.}\ }\textbf {\bibinfo {volume} {764}},\ \bibinfo {pages} {156} (\bibinfo {year} {2014})}\BibitemShut {NoStop}%
\bibitem [{\citenamefont {Brown}\ and\ \citenamefont {Altermatt}(1985)}]{Brown1985}%
  \BibitemOpen
  \bibfield  {author} {\bibinfo {author} {\bibfnamefont {I.~D.}\ \bibnamefont {Brown}}\ and\ \bibinfo {author} {\bibfnamefont {D.}~\bibnamefont {Altermatt}},\ }\bibfield  {title} {\bibinfo {title} {{Bond-Valence Parameters Obtained from a Systematic Analysis of the Inorganic Crystal Structure Database}},\ }\href@noop {} {\bibfield  {journal} {\bibinfo  {journal} {Acta Crystallogr.}\ }\textbf {\bibinfo {volume} {B41}},\ \bibinfo {pages} {244} (\bibinfo {year} {1985})}\BibitemShut {NoStop}%
\bibitem [{sup()}]{supplemental}%
  \BibitemOpen
  \href@noop {} {}\bibinfo {note} {See Supplemental Material at [URL will be inserted by publisher] for additional refinements and structural data from variable-temperature powder neutron diffraction.}\BibitemShut {Stop}%
\bibitem [{\citenamefont {MacChesney}\ \emph {et~al.}(1966)\citenamefont {MacChesney}, \citenamefont {Williams}, \citenamefont {Sherwood},\ and\ \citenamefont {Potter}}]{MacChesney1966}%
  \BibitemOpen
  \bibfield  {author} {\bibinfo {author} {\bibfnamefont {J.~B.}\ \bibnamefont {MacChesney}}, \bibinfo {author} {\bibfnamefont {H.~J.}\ \bibnamefont {Williams}}, \bibinfo {author} {\bibfnamefont {R.~C.}\ \bibnamefont {Sherwood}},\ and\ \bibinfo {author} {\bibfnamefont {J.~F.}\ \bibnamefont {Potter}},\ }\bibfield  {title} {\bibinfo {title} {{Magnetic properties of the terbium oxides at temperatures between 1.4° and 300°K}},\ }\href {https://doi.org/10.1063/1.1708501} {\bibfield  {journal} {\bibinfo  {journal} {J. Appl. Phys.}\ }\textbf {\bibinfo {volume} {37}},\ \bibinfo {pages} {1435} (\bibinfo {year} {1966})}\BibitemShut {NoStop}%
\bibitem [{\citenamefont {Ayant}\ \emph {et~al.}(1971)\citenamefont {Ayant}, \citenamefont {Belorizky},\ and\ \citenamefont {Tcheou}}]{Ayant1971}%
  \BibitemOpen
  \bibfield  {author} {\bibinfo {author} {\bibfnamefont {Y.}~\bibnamefont {Ayant}}, \bibinfo {author} {\bibfnamefont {E.}~\bibnamefont {Belorizky}},\ and\ \bibinfo {author} {\bibfnamefont {F.}~\bibnamefont {Tcheou}},\ }\bibfield  {title} {\bibinfo {title} {{Propri{\'{e}}t{\'{e}}s magn{\'{e}}tiques de \ch{Tb2O3} aux basses temp{\'{e}}ratures}},\ }\href@noop {} {\bibfield  {journal} {\bibinfo  {journal} {J. Phys. Colloq.}\ }\textbf {\bibinfo {volume} {32}},\ \bibinfo {pages} {1022} (\bibinfo {year} {1971})}\BibitemShut {NoStop}%
\bibitem [{\citenamefont {Hill}(1986)}]{Hill1986}%
  \BibitemOpen
  \bibfield  {author} {\bibinfo {author} {\bibfnamefont {R.~W.}\ \bibnamefont {Hill}},\ }\bibfield  {title} {\bibinfo {title} {{The specific heats of \ch{Tb2O3} and \ch{Tb4O7} between 0.5 and 22 K}},\ }\href@noop {} {\bibfield  {journal} {\bibinfo  {journal} {J. Phys. C Solid State Phys.}\ }\textbf {\bibinfo {volume} {19}},\ \bibinfo {pages} {67} (\bibinfo {year} {1986})}\BibitemShut {NoStop}%
\bibitem [{\citenamefont {Clark}\ \emph {et~al.}(2019)\citenamefont {Clark}, \citenamefont {Sala}, \citenamefont {Maharaj}, \citenamefont {Stone}, \citenamefont {Knight}, \citenamefont {Telling}, \citenamefont {Wang}, \citenamefont {Xu}, \citenamefont {Kim}, \citenamefont {Li}, \citenamefont {Cheong},\ and\ \citenamefont {Gaulin}}]{Clark2019}%
  \BibitemOpen
  \bibfield  {author} {\bibinfo {author} {\bibfnamefont {L.}~\bibnamefont {Clark}}, \bibinfo {author} {\bibfnamefont {G.}~\bibnamefont {Sala}}, \bibinfo {author} {\bibfnamefont {D.~D.}\ \bibnamefont {Maharaj}}, \bibinfo {author} {\bibfnamefont {M.~B.}\ \bibnamefont {Stone}}, \bibinfo {author} {\bibfnamefont {K.~S.}\ \bibnamefont {Knight}}, \bibinfo {author} {\bibfnamefont {M.~T.}\ \bibnamefont {Telling}}, \bibinfo {author} {\bibfnamefont {X.}~\bibnamefont {Wang}}, \bibinfo {author} {\bibfnamefont {X.}~\bibnamefont {Xu}}, \bibinfo {author} {\bibfnamefont {J.}~\bibnamefont {Kim}}, \bibinfo {author} {\bibfnamefont {Y.}~\bibnamefont {Li}}, \bibinfo {author} {\bibfnamefont {S.~W.}\ \bibnamefont {Cheong}},\ and\ \bibinfo {author} {\bibfnamefont {B.~D.}\ \bibnamefont {Gaulin}},\ }\bibfield  {title} {\bibinfo {title} {{Two-dimensional spin liquid behaviour in the triangular-honeycomb antiferromagnet \ch{TbInO3}}},\ }\href {https://doi.org/10.1038/s41567-018-0407-2} {\bibfield  {journal} {\bibinfo  {journal} {Nat.
  Phys.}\ }\textbf {\bibinfo {volume} {15}},\ \bibinfo {pages} {262} (\bibinfo {year} {2019})}\BibitemShut {NoStop}%
\bibitem [{\citenamefont {Stevens}(1952)}]{Stevens1952}%
  \BibitemOpen
  \bibfield  {author} {\bibinfo {author} {\bibfnamefont {K.~W.~H.}\ \bibnamefont {Stevens}},\ }\bibfield  {title} {\bibinfo {title} {{Matrix elements and operator equivalents connected with the magnetic properties of rare earth ions}},\ }\href {https://doi.org/10.1088/0370-1298/65/3/308} {\bibfield  {journal} {\bibinfo  {journal} {Proc. Phys. Soc. Sect. A}\ }\textbf {\bibinfo {volume} {65}},\ \bibinfo {pages} {209} (\bibinfo {year} {1952})}\BibitemShut {NoStop}%
\bibitem [{Note1()}]{Note1}%
  \BibitemOpen
  \bibinfo {note} {While these effective charges are significantly different from the formal charges on each ion, this is not unexpected given the significant covalency in the system, i.e.~the presence of \ch {BO3} molecular anions. In particular, one would expect the effective charge on the B atoms to be much lower than that of the more ionic Tb atoms, which is found to be the case.}\BibitemShut {Stop}%
\bibitem [{\citenamefont {Virtanen}\ \emph {et~al.}(2020)\citenamefont {Virtanen}, \citenamefont {Gommers}, \citenamefont {Oliphant}, \citenamefont {Haberland}, \citenamefont {Reddy}, \citenamefont {Cournapeau}, \citenamefont {Burovski}, \citenamefont {Peterson}, \citenamefont {Weckesser}, \citenamefont {Bright}, \citenamefont {{van der Walt}}, \citenamefont {Brett}, \citenamefont {Wilson}, \citenamefont {Millman}, \citenamefont {Mayorov}, \citenamefont {Nelson}, \citenamefont {Jones}, \citenamefont {Kern}, \citenamefont {Larson}, \citenamefont {Carey}, \citenamefont {Polat}, \citenamefont {Feng}, \citenamefont {Moore}, \citenamefont {{VanderPlas}}, \citenamefont {Laxalde}, \citenamefont {Perktold}, \citenamefont {Cimrman}, \citenamefont {Henriksen}, \citenamefont {Quintero}, \citenamefont {Harris}, \citenamefont {Archibald}, \citenamefont {Ribeiro}, \citenamefont {Pedregosa}, \citenamefont {{van Mulbregt}},\ and\ \citenamefont {{SciPy 1.0 Contributors}}}]{scipy}%
  \BibitemOpen
  \bibfield  {author} {\bibinfo {author} {\bibfnamefont {P.}~\bibnamefont {Virtanen}}, \bibinfo {author} {\bibfnamefont {R.}~\bibnamefont {Gommers}}, \bibinfo {author} {\bibfnamefont {T.~E.}\ \bibnamefont {Oliphant}}, \bibinfo {author} {\bibfnamefont {M.}~\bibnamefont {Haberland}}, \bibinfo {author} {\bibfnamefont {T.}~\bibnamefont {Reddy}}, \bibinfo {author} {\bibfnamefont {D.}~\bibnamefont {Cournapeau}}, \bibinfo {author} {\bibfnamefont {E.}~\bibnamefont {Burovski}}, \bibinfo {author} {\bibfnamefont {P.}~\bibnamefont {Peterson}}, \bibinfo {author} {\bibfnamefont {W.}~\bibnamefont {Weckesser}}, \bibinfo {author} {\bibfnamefont {J.}~\bibnamefont {Bright}}, \bibinfo {author} {\bibfnamefont {S.~J.}\ \bibnamefont {{van der Walt}}}, \bibinfo {author} {\bibfnamefont {M.}~\bibnamefont {Brett}}, \bibinfo {author} {\bibfnamefont {J.}~\bibnamefont {Wilson}}, \bibinfo {author} {\bibfnamefont {K.~J.}\ \bibnamefont {Millman}}, \bibinfo {author} {\bibfnamefont {N.}~\bibnamefont {Mayorov}}, \bibinfo {author} {\bibfnamefont
  {A.~R.~J.}\ \bibnamefont {Nelson}}, \bibinfo {author} {\bibfnamefont {E.}~\bibnamefont {Jones}}, \bibinfo {author} {\bibfnamefont {R.}~\bibnamefont {Kern}}, \bibinfo {author} {\bibfnamefont {E.}~\bibnamefont {Larson}}, \bibinfo {author} {\bibfnamefont {C.~J.}\ \bibnamefont {Carey}}, \bibinfo {author} {\bibfnamefont {{\.I}.}~\bibnamefont {Polat}}, \bibinfo {author} {\bibfnamefont {Y.}~\bibnamefont {Feng}}, \bibinfo {author} {\bibfnamefont {E.~W.}\ \bibnamefont {Moore}}, \bibinfo {author} {\bibfnamefont {J.}~\bibnamefont {{VanderPlas}}}, \bibinfo {author} {\bibfnamefont {D.}~\bibnamefont {Laxalde}}, \bibinfo {author} {\bibfnamefont {J.}~\bibnamefont {Perktold}}, \bibinfo {author} {\bibfnamefont {R.}~\bibnamefont {Cimrman}}, \bibinfo {author} {\bibfnamefont {I.}~\bibnamefont {Henriksen}}, \bibinfo {author} {\bibfnamefont {E.~A.}\ \bibnamefont {Quintero}}, \bibinfo {author} {\bibfnamefont {C.~R.}\ \bibnamefont {Harris}}, \bibinfo {author} {\bibfnamefont {A.~M.}\ \bibnamefont {Archibald}}, \bibinfo {author}
  {\bibfnamefont {A.~H.}\ \bibnamefont {Ribeiro}}, \bibinfo {author} {\bibfnamefont {F.}~\bibnamefont {Pedregosa}}, \bibinfo {author} {\bibfnamefont {P.}~\bibnamefont {{van Mulbregt}}},\ and\ \bibinfo {author} {\bibnamefont {{SciPy 1.0 Contributors}}},\ }\bibfield  {title} {\bibinfo {title} {{{SciPy} 1.0: Fundamental Algorithms for Scientific Computing in Python}},\ }\href {https://doi.org/10.1038/s41592-019-0686-2} {\bibfield  {journal} {\bibinfo  {journal} {Nature Methods}\ }\textbf {\bibinfo {volume} {17}},\ \bibinfo {pages} {261} (\bibinfo {year} {2020})}\BibitemShut {NoStop}%
\bibitem [{\citenamefont {Jiang}\ \emph {et~al.}(2022)\citenamefont {Jiang}, \citenamefont {Yang}, \citenamefont {Gao}, \citenamefont {Wan}, \citenamefont {Zhu}, \citenamefont {Shiroka}, \citenamefont {Chen}, \citenamefont {Wu}, \citenamefont {Li}, \citenamefont {Jiao}, \citenamefont {Chen}, \citenamefont {Bao}, \citenamefont {Tian},\ and\ \citenamefont {Shu}}]{Jiang2022}%
  \BibitemOpen
  \bibfield  {author} {\bibinfo {author} {\bibfnamefont {C.~Y.}\ \bibnamefont {Jiang}}, \bibinfo {author} {\bibfnamefont {Y.~X.}\ \bibnamefont {Yang}}, \bibinfo {author} {\bibfnamefont {Y.~X.}\ \bibnamefont {Gao}}, \bibinfo {author} {\bibfnamefont {Z.~T.}\ \bibnamefont {Wan}}, \bibinfo {author} {\bibfnamefont {Z.~H.}\ \bibnamefont {Zhu}}, \bibinfo {author} {\bibfnamefont {T.}~\bibnamefont {Shiroka}}, \bibinfo {author} {\bibfnamefont {C.~S.}\ \bibnamefont {Chen}}, \bibinfo {author} {\bibfnamefont {Q.}~\bibnamefont {Wu}}, \bibinfo {author} {\bibfnamefont {X.}~\bibnamefont {Li}}, \bibinfo {author} {\bibfnamefont {J.~C.}\ \bibnamefont {Jiao}}, \bibinfo {author} {\bibfnamefont {K.~W.}\ \bibnamefont {Chen}}, \bibinfo {author} {\bibfnamefont {Y.}~\bibnamefont {Bao}}, \bibinfo {author} {\bibfnamefont {Z.~M.}\ \bibnamefont {Tian}},\ and\ \bibinfo {author} {\bibfnamefont {L.}~\bibnamefont {Shu}},\ }\bibfield  {title} {\bibinfo {title} {Spin excitations in the quantum dipolar magnet
  $\mathrm{Yb}{({\mathrm{BaBO}}_{3})}_{3}$},\ }\href {https://doi.org/10.1103/PhysRevB.106.014409} {\bibfield  {journal} {\bibinfo  {journal} {Phys. Rev. B}\ }\textbf {\bibinfo {volume} {106}},\ \bibinfo {pages} {014409} (\bibinfo {year} {2022})}\BibitemShut {NoStop}%
\bibitem [{\citenamefont {Mukherjee}\ \emph {et~al.}(2017)\citenamefont {Mukherjee}, \citenamefont {{Sackville Hamilton}}, \citenamefont {Glass},\ and\ \citenamefont {Dutton}}]{Mukherjee2017a}%
  \BibitemOpen
  \bibfield  {author} {\bibinfo {author} {\bibfnamefont {P.}~\bibnamefont {Mukherjee}}, \bibinfo {author} {\bibfnamefont {A.~C.}\ \bibnamefont {{Sackville Hamilton}}}, \bibinfo {author} {\bibfnamefont {H.~F.~J.}\ \bibnamefont {Glass}},\ and\ \bibinfo {author} {\bibfnamefont {S.~E.}\ \bibnamefont {Dutton}},\ }\bibfield  {title} {\bibinfo {title} {{Sensitivity of magnetic properties to chemical pressure in lanthanide garnets \ch{\textit{Ln}3\textit{A}2\textit{X}3O12}, \textit{Ln} = Gd, Tb, Dy, Ho, \textit{A} = Ga, Sc, In, Te, \textit{X} = Ga, Al, Li}},\ }\href@noop {} {\bibfield  {journal} {\bibinfo  {journal} {J. Phys. Condens. Matter}\ }\textbf {\bibinfo {volume} {29}},\ \bibinfo {pages} {405808} (\bibinfo {year} {2017})}\BibitemShut {NoStop}%
\bibitem [{\citenamefont {Hammann}\ and\ \citenamefont {Manneville}(1973)}]{Hammann1973}%
  \BibitemOpen
  \bibfield  {author} {\bibinfo {author} {\bibfnamefont {J.}~\bibnamefont {Hammann}}\ and\ \bibinfo {author} {\bibfnamefont {P.}~\bibnamefont {Manneville}},\ }\bibfield  {title} {\bibinfo {title} {{Ordre magn{\'{e}}tique {\'{e}}lectronique induit par les interactions hyperfines dans les grenats de gallium-holmium et de gallium-terbium}},\ }\href@noop {} {\bibfield  {journal} {\bibinfo  {journal} {J. Phys. Fr.}\ }\textbf {\bibinfo {volume} {34}},\ \bibinfo {pages} {615} (\bibinfo {year} {1973})}\BibitemShut {NoStop}%
\bibitem [{\citenamefont {Mirebeau}\ \emph {et~al.}(2005)\citenamefont {Mirebeau}, \citenamefont {Apetrei}, \citenamefont {Rodr{\'{i}}guez-Carvajal}, \citenamefont {Bonville}, \citenamefont {Forget}, \citenamefont {Colson}, \citenamefont {Glazkov}, \citenamefont {Sanchez}, \citenamefont {Isnard},\ and\ \citenamefont {Suard}}]{Mirebeau2005}%
  \BibitemOpen
  \bibfield  {author} {\bibinfo {author} {\bibfnamefont {I.}~\bibnamefont {Mirebeau}}, \bibinfo {author} {\bibfnamefont {A.}~\bibnamefont {Apetrei}}, \bibinfo {author} {\bibfnamefont {J.}~\bibnamefont {Rodr{\'{i}}guez-Carvajal}}, \bibinfo {author} {\bibfnamefont {P.}~\bibnamefont {Bonville}}, \bibinfo {author} {\bibfnamefont {A.}~\bibnamefont {Forget}}, \bibinfo {author} {\bibfnamefont {D.}~\bibnamefont {Colson}}, \bibinfo {author} {\bibfnamefont {V.}~\bibnamefont {Glazkov}}, \bibinfo {author} {\bibfnamefont {J.~P.}\ \bibnamefont {Sanchez}}, \bibinfo {author} {\bibfnamefont {O.}~\bibnamefont {Isnard}},\ and\ \bibinfo {author} {\bibfnamefont {E.}~\bibnamefont {Suard}},\ }\bibfield  {title} {\bibinfo {title} {{Ordered spin ice state and magnetic fluctuations in \ch{Tb2Sn2O7}}},\ }\href {https://doi.org/10.1103/PhysRevLett.94.246402} {\bibfield  {journal} {\bibinfo  {journal} {Phys. Rev. Lett.}\ }\textbf {\bibinfo {volume} {94}},\ \bibinfo {pages} {246402} (\bibinfo {year} {2005})}\BibitemShut {NoStop}%
\bibitem [{\citenamefont {Pan}\ \emph {et~al.}(2021)\citenamefont {Pan}, \citenamefont {Ni}, \citenamefont {He}, \citenamefont {Yu}, \citenamefont {Xu},\ and\ \citenamefont {Li}}]{Pan2021}%
  \BibitemOpen
  \bibfield  {author} {\bibinfo {author} {\bibfnamefont {B.~L.}\ \bibnamefont {Pan}}, \bibinfo {author} {\bibfnamefont {J.~M.}\ \bibnamefont {Ni}}, \bibinfo {author} {\bibfnamefont {L.~P.}\ \bibnamefont {He}}, \bibinfo {author} {\bibfnamefont {Y.~J.}\ \bibnamefont {Yu}}, \bibinfo {author} {\bibfnamefont {Y.}~\bibnamefont {Xu}},\ and\ \bibinfo {author} {\bibfnamefont {S.~Y.}\ \bibnamefont {Li}},\ }\bibfield  {title} {\bibinfo {title} {Specific heat and thermal conductivity of the triangular-lattice rare-earth material $\mathrm{KBaYb}{({\mathrm{BO}}_{3})}_{2}$ at ultralow temperature},\ }\href {https://doi.org/10.1103/PhysRevB.103.104412} {\bibfield  {journal} {\bibinfo  {journal} {Phys. Rev. B}\ }\textbf {\bibinfo {volume} {103}},\ \bibinfo {pages} {104412} (\bibinfo {year} {2021})}\BibitemShut {NoStop}%
\bibitem [{\citenamefont {Mustonen}\ \emph {et~al.}(2018)\citenamefont {Mustonen}, \citenamefont {Vasala}, \citenamefont {Sadrollahi}, \citenamefont {Schmidt}, \citenamefont {Baines}, \citenamefont {Walker}, \citenamefont {Terasaki}, \citenamefont {Litterst}, \citenamefont {Baggio-Saitovitch},\ and\ \citenamefont {Karppinen}}]{Mustonen2018}%
  \BibitemOpen
  \bibfield  {author} {\bibinfo {author} {\bibfnamefont {O.}~\bibnamefont {Mustonen}}, \bibinfo {author} {\bibfnamefont {S.}~\bibnamefont {Vasala}}, \bibinfo {author} {\bibfnamefont {E.}~\bibnamefont {Sadrollahi}}, \bibinfo {author} {\bibfnamefont {K.}~\bibnamefont {Schmidt}}, \bibinfo {author} {\bibfnamefont {C.}~\bibnamefont {Baines}}, \bibinfo {author} {\bibfnamefont {H.}~\bibnamefont {Walker}}, \bibinfo {author} {\bibfnamefont {I.}~\bibnamefont {Terasaki}}, \bibinfo {author} {\bibfnamefont {F.}~\bibnamefont {Litterst}}, \bibinfo {author} {\bibfnamefont {E.}~\bibnamefont {Baggio-Saitovitch}},\ and\ \bibinfo {author} {\bibfnamefont {M.}~\bibnamefont {Karppinen}},\ }\bibfield  {title} {\bibinfo {title} {Spin-liquid-like state in a spin-$\frac{1}{2}$ square-lattice antiferromagnet perovskite induced by d$^{10}$–d$^0$ cation mixing},\ }\href {https://doi.org/10.1038/s41467-018-03435-1} {\bibfield  {journal} {\bibinfo  {journal} {Nat. Commun.}\ }\textbf {\bibinfo {volume} {9}},\ \bibinfo {pages} {1085} (\bibinfo {year}
  {2018})}\BibitemShut {NoStop}%
\bibitem [{\citenamefont {Khatua}\ \emph {et~al.}(2022)\citenamefont {Khatua}, \citenamefont {Orain}, \citenamefont {Ozarowski}, \citenamefont {Sethupathi}, \citenamefont {Rao}, \citenamefont {Zorko},\ and\ \citenamefont {Khuntia}}]{Khatua2022}%
  \BibitemOpen
  \bibfield  {author} {\bibinfo {author} {\bibfnamefont {J.}~\bibnamefont {Khatua}}, \bibinfo {author} {\bibfnamefont {J.}~\bibnamefont {Orain}}, \bibinfo {author} {\bibfnamefont {A.}~\bibnamefont {Ozarowski}}, \bibinfo {author} {\bibfnamefont {K.}~\bibnamefont {Sethupathi}}, \bibinfo {author} {\bibfnamefont {M.~S.}\ \bibnamefont {Rao}}, \bibinfo {author} {\bibfnamefont {A.}~\bibnamefont {Zorko}},\ and\ \bibinfo {author} {\bibfnamefont {P.}~\bibnamefont {Khuntia}},\ }\bibfield  {title} {\bibinfo {title} {Signature of a randomness-driven spin-liquid state in a frustrated magnet},\ }\href@noop {} {\bibfield  {journal} {\bibinfo  {journal} {Commun. Phys.}\ }\textbf {\bibinfo {volume} {5}},\ \bibinfo {pages} {99} (\bibinfo {year} {2022})}\BibitemShut {NoStop}%
\bibitem [{\citenamefont {Nuccio}\ \emph {et~al.}(2014)\citenamefont {Nuccio}, \citenamefont {Schulz},\ and\ \citenamefont {Drew}}]{Nuccio2014}%
  \BibitemOpen
  \bibfield  {author} {\bibinfo {author} {\bibfnamefont {L.}~\bibnamefont {Nuccio}}, \bibinfo {author} {\bibfnamefont {L.}~\bibnamefont {Schulz}},\ and\ \bibinfo {author} {\bibfnamefont {A.~J.}\ \bibnamefont {Drew}},\ }\bibfield  {title} {\bibinfo {title} {Muon spin spectroscopy: magnetism, soft matter and the bridge between the two},\ }\href {https://doi.org/10.1088/0022-3727/47/47/473001} {\bibfield  {journal} {\bibinfo  {journal} {Journal of Physics D: Applied Physics}\ }\textbf {\bibinfo {volume} {47}},\ \bibinfo {pages} {473001} (\bibinfo {year} {2014})}\BibitemShut {NoStop}%
\end{thebibliography}
\end{document}